\RequirePackage{xr-hyper}							
\documentclass[]{volcanica-template}
\nonstopmode									
\titleformat{\paragraph}[hang]{\normalfont\bfseries}{}{0pt}{}		
\titlespacing*{\paragraph}{0pt}{1.5ex}{0.5ex}
\usepackage{placeins}							

\newcommand{\gE}{g^{\mathrm{E}}}
\newcommand{\ESPhi}{openCOSMO-RS-ES-Phi}
\newcommand{\oPhi}{openCOSMO-RS-Phi}
\newcommand{\mycomment}[1]{}
\makeatletter										
\@addtofilelist{SI.tex}
\@addtofilelist{SI.aux}\IfFileExists{SI.aux}{\typeout{(SI.aux)}}{\typeout{No file SI.aux.}}
\makeatother

\newcommand{\ArticleType}{Research article}			
\newcommand{\Title}{An Electrolyte Equation of State Based on openCOSMO-RS: From Aqueous Solutions to Non-Aqueous Systems} 	
\newcommand{\shortTitle}{EOS Electrolytes} 		        
\newcommand{\Author}{Raudszus and M\"uller\xspace}    	

\author[{{\affiliation{1}}}] 				
{\orcidaffil{0009-0006-6547-4786}~			
Leander Raudszus} 					        	
\author[{{\affiliation{1}}}] 				
{\orcidaffil{0000-0003-1684-6994}~			
Simon Müller\Email{simon.mueller@tuhh.de}} 		

\affil[{{\affiliation{1}}}]{					
Institute of Thermal Separation Processes, Hamburg University of Technology, Hamburg, Germany
}

\begin{document}
\nolinenumbers
\FrontMatter{
Predictive COSMO-RS-based electrolyte models such as COSMO-RS-ES describe electrolyte solutions from quantum-chemical information, but treat the liquid as incompressible and give no access to densities or pressure effects.
This work presents \ESPhi{}, to our knowledge the first open-source implementation of a predictive electrolyte equation of state based on openCOSMO-RS.
It combines the COSMO-RS-based equation of state \oPhi{} with the ion interaction energies and the long-range term of COSMO-RS-ES, which is first refitted on the same simplified short range.
With this refit, a dielectric decrement improves the mean ionic activity coefficients (MIAC), whereas a Born term counts the solvation of the ions a second time and is therefore left out.
The equation of state then adds only volume, dispersion and temperature parameters of the ions, fitted to aqueous MIAC and densities.
\ESPhi{} reproduces COSMO-RS-ES in dilute aqueous solutions, improves on it in concentrated ones and also yields solution densities.
Extrapolating from aqueous systems, it predicts the MIAC in water + cosolvent mixtures as accurately as the fitted COSMO-RS-ES, as well as the MIAC and bubble points of a benchmark database of water--alcohol--salt systems.
Salting-out of organic components is not described as accurately, and salts come out too soluble in organic solvents.
Overall, the interaction model developed for COSMO-RS-ES carries over to a large extent to an equation of state, which makes \ESPhi{} a sound base model for further development, with ion transfer between solvents as a clear target for fitting to non-aqueous data.
}
[]%
{					
{COSMO-RS}{Open-source}{Parametrization}{ORCA}{Equation of State}	
}

\hypertarget{introduction}{
\section{Introduction}\label{introduction}}

Reliable predictions of the thermodynamic properties of mixed solvents, and of electrolyte solutions in particular, are of considerable importance to both industry and academia. Dissolved salts change the phase behavior, the volatility and the crystallization behavior of a solvent mixture substantially, and they do so across a range of compositions, temperatures and pressures that experiments alone can rarely cover. Predictive models therefore allow large numbers of candidate solvents to be screened and new processes, such as crystallization, extraction or salt-assisted distillation, to be designed with only a limited number of measurements.

A number of predictive approaches are available for incompressible neutral and electrolyte systems.
Activity coefficient models, formulated as an excess Gibbs energy $\gE$, describe the composition-dependent deviation from ideal behavior at a given temperature and pressure. The most prominent model classes, NRTL, UNIQUAC and COSMO-based models, differ in their predictive power and in the number of parameters they require.
Their electrolyte variants build on the Debye--H\"uckel (DH) term for the long-range electrostatic interactions between ions \citep{Hist12Debye1, kontogeorgisDebyeHuckelTheoryIts2018}, which is sometimes applied in its Pitzer--Debye--H\"uckel (PDH) form and often combined with a Born term that accounts for the transfer of ions between media of different permittivity.

The various electrolyte NRTL formulations share the long-range term and differ mainly in their short-range term \citep{chenLocalCompositionModel1982,chenLocalCompositionModel1986}, and a Born term was added for mixed solvents \citep{chenGeneralizedElectrolyteNRTLModel2004}.
Nevertheless, eNRTL remains a correlative model, since its binary interaction parameters have to be regressed for every ion pair or salt-solvent combination.

Electrolyte UNIQUAC models have likewise been applied successfully to correlate phase equilibria in liquid electrolyte systems \citep{sanderCalculationVapourliquidEquilibria1986,thomsenModelingVaporLiquid1999}. The LIQUAC model evaluates its long-range term with the permittivity of the salt-free solvent mixture \citep{liGEModelSingle1994}, whereas \citet{wangComputationDielectricConstants2001} introduced a permittivity that also depends on the salt concentration in a mixed-solvent electrolyte model.

UNIFAC-based approaches reach a higher, semi-predictive level by transferring interaction parameters between chemically related species through group contributions.
The resulting LIFAC and AIOMFAC models correlate and predict phase equilibria and mean ionic activity coefficients (MIACs) \citep{yanPredictionVaporLiquid1999,zuendThermodynamicModelMixed2008,mohsRevisedLIQUACLIFAC2013}, but they remain restricted to the groups covered by their parameter tables.

Models based on COSMO theory, in contrast, compute the short-range contribution from screening charge densities ($\sigma$-profiles) obtained by quantum chemistry, so that the behavior of a mixture follows from pure-component information and a small set of globally fitted parameters \citep{klamtConductorlikeScreeningModel1995,klamtRefinementParametrizationCOSMORS1998, linPrioriPhaseEquilibrium2002,gerlachOpenSourceCOSMORS2022,mullerPredictingSolvationFree2025}.
The open-source implementation openCOSMO-RS \citep{gerlachOpenSourceCOSMORS2022} has since been extended by a dispersion contribution \citep{grigorashComprehensiveApproachIncorporating2025,grigorashComprehensiveApproachIncorporating2026} and applied to the solid--liquid equilibria of eutectic systems \citep{ahmedFastAccuratePrediction2026}.

Early electrolyte extensions combined this description with the PDH term \citep{hsiehPredictiveModelExcess2011,wangExtensionCOSMOSACSolvation2011,toureCOSMORSPDHSNewPredictive2014}, in one case together with scaled ion-specific hydrogen-bonding energies \citep{ingramExtensionCOSMORSMonoatomic2012}.
Its successor from the same group, the COSMO-RS-ES model \citep{gerlachDevelopmentCOSMORSBased2018}, replaced the scaled hydrogen-bonding energies by general empirical interaction-energy equations, again combined with the PDH term.
M\"uller et al.\ evaluated this model on solid-liquid equilibria and refined it using Gibbs free energies of ion transfer \citep{mullerEvaluationRefinementNovel2019a} and for systems of low permittivity \citep{mullerCalculationThermodynamicEquilibria2020}.
COSMO-RS-ES also predicts solvation free energies of ions in neutral solvents with an accuracy comparable to that of a cluster-continuum approach \citep{krogerPredictionSolvationFree2020}, and curated reference data for the solvation free energies of anions in several solvents have recently become available \citep{nevolianisSolvationFreeEnergies2025}.
Gonz\'alez de Castilla et al.\ subsequently made the long-range term Gibbs-Duhem consistent, removed the conventionally fixed distance of closest approach, and corrected over- and underscreening at low and variable relative permittivity \citep{gonzalezdecastillaAnalogyRestrictedPrimitive2021,gonzalezdecastillaAnalogyRestrictedPrimitive2022}, before revising the short-range term to obtain COSMO-RS-ES+ \citep{gonzalezdecastillaRevisionExtensionCOSMORS2025}.
These models predict SLE, LLE and MIACs of aqueous and mixed-solvent electrolytes and, most recently, of ionic liquids up to the pure ionic liquid state \citep{marquesImpactExtendedLongrange2023,gonzalezdecastillaRevisionExtensionCOSMORS2025}. In addition, COSMO-RS-ES, fitted at 298.15~K, has been applied to the cesium nitrate--water system over a wider temperature range \citep{arradThermodynamicModelingUsing2024}.

Since activity coefficient models treat the liquid as incompressible, its molar volume is not an independent variable and has to be supplied externally.
Densities, partial molar volumes, compressibilities and the pressure dependence of phase equilibria and solubilities therefore remain out of reach.
Access to these properties requires an equation of state.

The first widely used electrolyte equation of state was proposed by \citet{furstRepresentationExcessProperties1993}. It combines a cubic equation of state with a mean-spherical-approximation (MSA) term for the long-range interactions of the ions and an ion-specific short-range term.
Subsequent models retain a physical reference term and an ionic term based on DH or MSA theory, and some add a Born term for the solvation of the ions \citep{kontogeorgisReviewElectrolyteEquations2022}.
Electrolyte members of the SAFT family include SAFT-VRE \citep{galindoSAFTVREPhaseBehavior1999b,schreckenbergModellingThermodynamicSolvation2014}, ePC-SAFT \citep{camerettiModelingAqueousElectrolyte2005,heldModelingAqueousElectrolyte2008,heldEPCSAFTRevised2014,bulowEPCSAFTAdvancedPart2021b,bulowEPCSAFTAdvancedPart2021a}, SAFT-VRE Mie \citep{eriksenDevelopmentIntermolecularPotential2016} and eSAFT-VR Mie \citep{selamThermodynamicModelStrong2018}, and the related cubic-plus-association (CPA) equation of state has been extended in the same way to eCPA \citep{maribo-mogensenElectrolyteCPAEquation2015}.
These models have been reviewed by \citet{kontogeorgisReviewElectrolyteEquations2022}, with a focus on cubic and CPA equations of state, and by \citet{novakMixedSolventElectrolyte2023}, with a focus on mixed solvents.
Two issues recur throughout this literature. The first is the relative static permittivity. Its decrement with salt concentration and, in a volume-explicit framework, its density dependence enter both the DH and the Born term, and predictions are highly sensitive to how they are described \citep{walkerImportanceRelativeStatic2022,silvaHowAccountConcentration2023,olsenComparisonsEquationState2023}.
The second is the set of ion parameters, which are assigned either per ion or per salt and have to be regressed from experimental data \citep{yangSystematicEvaluationParameterization2023}.
This limits predictivity, since every new ion must be parametrized against experimental data before it can be described.
Beyond water, \citet{novakExtensionESAFTVRMie2023} fitted the ion parameters of eSAFT-VR Mie to aqueous MIACs, individual ion activity coefficients and densities and, using a mole-fraction mixing rule for the permittivity, predicted the MIACs of single salts in methanol and ethanol without further parameters.
Volumetric properties are thus well established in this branch, but the ion parameters have so far been regressed mostly from aqueous data, and the water-poor regime, where models still require reparameterization or solvent-dependent ion parameters, remains the least developed \citep{heldThermodynamicGEModels2020}.

The two branches can nevertheless be reconciled. Lattice-fluid theories admit vacant sites, or holes, which turns the fluid into a pseudo-mixture of the real components and a hole pseudo-component and makes the volume an independent variable \citep{sanchezElementaryMolecularTheory1976,panayiotouStatisticalThermodynamicsRMer1982}. While not implementing lattice-fluid theory, \citet{dep.soaresPairwiseSurfaceContact2019} applied the idea of adding holes as a pseudo component to COSMO-SAC and obtained the pairwise surface-contact equation of
state COSMO-SAC-Phi, whose four pure-component parameters are fitted to saturation pressures and saturated liquid molar volumes, and \citet{zini_improved_2021} later refined its dispersive contribution.

\citet{markgrafExtensionOpenCOSMORSFull2026} implemented the model into openCOSMO-RS as the Helmholtz-explicit equation of state openCOSMO-RS-Phi and provided an open parameter set for some 1800 substances, benchmarked on a large set of systems without binary interaction parameters.

The aim of this work is therefore to combine the equation of state openCOSMO-RS-Phi of \citet{markgrafExtensionOpenCOSMORSFull2026} with the predictive COSMO-based electrolyte models discussed above, by incorporating the ion interaction energies and the long-range term of COSMO-RS-ES into this equation of state, which yields \ESPhi{}.

\hypertarget{methodology}{%
\section{Methodology}\label{methodology}}

\subsection{Models}\label{sec:models}

COSMO-RS-ES treats the liquid as incompressible and serves as the reference activity coefficient model. Previous work \citep{markgrafExtensionOpenCOSMORSFull2026} showed that for mixtures the equation of state behaves like the underlying activity coefficient model, so the influence of the dielectric decrement and the Born term is studied with COSMO-RS-ES.
\ESPhi{} is the full equation of state and is applied to densities, to the temperature dependence of the aqueous MIAC and to the prediction of non-aqueous systems.
Both models share the $\sigma$-profiles, the short range of the neutral species with the global parameters of \citet{markgrafExtensionOpenCOSMORSFull2026}, the ion interaction energies of COSMO-RS-ES, the ME-PDH term for fully dissociated ions \citep{gonzalezdecastillaAnalogyRestrictedPrimitive2022,gonzalezdecastillaRevisionExtensionCOSMORS2025} and the reference state of the ions, infinite dilution in the salt-free solvent.
The $\sigma$-profiles of the ions are computed with the same ORCA workflow and settings as those of the neutral species \citep{mullerPredictingSolvationFree2025,markgrafExtensionOpenCOSMORSFull2026}.

To keep \oPhi{} as close as possible to COSMO-SAC-Phi \citep{dep.soaresPairwiseSurfaceContact2019}, its implementation uses a simplified short range, in which the electrostatic misfit keeps its basic form without the correlation term and the hydrogen bond has a single strength without temperature dependence \citep{markgrafExtensionOpenCOSMORSFull2026}.
Already for neutral systems, this simplified model is slightly less accurate than other openCOSMO-RS implementations with the full short range \citep{gerlachOpenSourceCOSMORS2022,mullerPredictingSolvationFree2025}.
Both models of this work build on the same simplified basis.

\subsubsection{COSMO-RS-ES}\label{sec:methods-es}

The COSMO-RS-ES version used in this work is based on the openCOSMO-RS implementation used in \ESPhi{}, and Table~\ref{tab:es-equations} collects its equations.

\begin{table}[!htb]
\centering
\caption{Equations of COSMO-RS-ES.
$\sigma_J$: screening charge density of segment type $J$, $\sigma_\mathrm{don}$ and $\sigma_\mathrm{acc}$ for a donor and an acceptor segment; $A_J^i$: area of segment type $J$ on species $i$; $\Theta_J$: share of segment type $J$ in the mixture; $V_i$: COSMO volume; $n$: total amount of substance; $M$, $\rho$, $\varepsilon_\mathrm{r}$: molar mass, density and permittivity of the mixture; $M_\mathrm{salt}$, $\nu$: molar mass and number of ions of the salt.
The energies $E^\mathrm{ion}_{JK}$ are given in Table~\ref{SI-tab:es-contacts} of the Supporting Information.}
\label{tab:es-equations}
\small
\begin{tabular}{p{0.17\textwidth}p{0.78\textwidth}}
\hline
Part & Equation \\
\hline
Activity coefficient & $\ln\gamma_i=\ln\gamma_i^\mathrm{C}+\ln\gamma_i^\mathrm{R}+\ln\gamma_i^\mathrm{LR}$, each part minus its value in the reference state of $i$, the pure liquid for a solvent and infinite dilution in the salt-free solvent for an ion \\[3pt]
Combinatorial & $\ln\gamma_i^\mathrm{C}=\ln\dfrac{V_i}{\sum_jx_jV_j}+1-\dfrac{V_i}{\sum_jx_jV_j}$ \\[8pt]
Residual & $\ln\gamma_i^\mathrm{R}=\displaystyle\sum_J\frac{A_J^i}{a_\mathrm{eff}}\ln\Gamma_J$, \quad $\Gamma_J=\Big[\displaystyle\sum_K\Theta_K\Gamma_K\exp\left(-\frac{E_{JK}}{RT}\right)\Big]^{-1}$ \\[8pt]
Neutral contacts & $E_{JK}=E^\mathrm{mf}_{JK}+E^\mathrm{hb}_{JK}$, \quad $E^\mathrm{mf}_{JK}=\dfrac{a_\mathrm{eff}\alpha}{2}\left(\sigma_J+\sigma_K\right)^2$, \newline
$E^\mathrm{hb}_{JK}=a_\mathrm{eff}c_\mathrm{hb}\min\left(0,\sigma_\mathrm{don}+\sigma_\mathrm{hb}\right)\max\left(0,\sigma_\mathrm{acc}-\sigma_\mathrm{hb}\right)$, with the donor segment on H or N and the acceptor segment on an atom other than H \\[3pt]
Contacts with ions & $E_{JK}=A_kE^\mathrm{mf}_{JK}+E^\mathrm{ion}_{JK}$ \\[3pt]
Long range$^a$ & $\ln\gamma_i^\mathrm{LR}=-A_\phi\left\{\dfrac{2z_i^2}{b_\mathrm{m}}L+\dfrac{z_i^2I_x^{1/2}-2I_x^{3/2}M_i/M}{1+b_\mathrm{m}I_x^{1/2}}-\dfrac{2I_x^{3/2}}{1+b_\mathrm{m}I_x^{1/2}}\left[\left(1+\dfrac{3b_1}{b_\mathrm{m}}\right)\varepsilon_i'-\rho_i'\right]-\dfrac{2I_xL}{b_\mathrm{m}}\left(2-\dfrac{3b_1}{b_\mathrm{m}}\right)\varepsilon_i'\right\}$ \\[8pt]
 & $L=\ln\left(1+b_\mathrm{m}I_x^{1/2}\right)$, \quad $I_x=\tfrac12\sum_jx_jz_j^2$, \quad $\varepsilon_i'=\dfrac{n}{\varepsilon_\mathrm{r}}\dfrac{\partial\varepsilon_\mathrm{r}}{\partial n_i}$, \quad $\rho_i'=\dfrac{n}{\rho}\dfrac{\partial\rho}{\partial n_i}$ \\[8pt]
 & $A_\phi=\dfrac13\left(\dfrac{2\pi N_\mathrm{A}\rho}{M}\right)^{1/2}\lambda_\mathrm{B}^{3/2}$, \quad $b=a\left(\dfrac{8\pi N_\mathrm{A}\rho\lambda_\mathrm{B}}{M}\right)^{1/2}$, \quad $\lambda_\mathrm{B}=\dfrac{e^2}{4\pi\varepsilon_0\varepsilon_\mathrm{r}k_\mathrm{B}T}$ \\[8pt]
 & $b_\mathrm{m}=\omega_0b+b_1$, \quad $b_1=\left(\dfrac{\omega_1\lambda_\mathrm{B}}{a}\right)^{3/2}b$ \\[8pt]
Closest approach & $a=F_\mathrm{M}\sum_{i\in\mathrm{M}}r_i+F_\mathrm{P}\sum_{i\in\mathrm{P}}\Psi_ir_i$, summed over cation and anion, with M the monoatomic and P the polyatomic ions and $\Psi_i$ the sphericity \\[3pt]
Mixture properties & $\rho=\sum_i\phi_i\rho_i$, \quad $\varepsilon_\mathrm{r}=\sum_i\phi_i\varepsilon_{\mathrm{r},i}$, \quad $\phi_i=\dfrac{x_iM_i/\rho_i}{\sum_jx_jM_j/\rho_j}$ \\[8pt]
Ions as molten salt & $\rho_i=973.35\left(\dfrac{M_\mathrm{salt}}{\mathrm{g\,mol^{-1}}}\right)^{0.1536}$~kg~m$^{-3}$ for salts of monoatomic cations and of ammonium, otherwise from the correlation for ionic liquids of \citet[Eq.~6.50]{gonzalezdecastillaRevisionExtensionCOSMORS2025}, \newline
$\varepsilon_{\mathrm{r},i}=1+\varepsilon_A\left(\dfrac{298.15~\mathrm{K}}{T}\right)^{\varepsilon_B}$, \quad $M_i=\dfrac{M_\mathrm{salt}}{\nu}$ \\[3pt]
\hline
\multicolumn{2}{p{0.95\textwidth}}{\footnotesize $^a$ \citet{gonzalezdecastillaAnalogyRestrictedPrimitive2022,gonzalezdecastillaRevisionExtensionCOSMORS2025}.} \\
\hline
\end{tabular}
\end{table}

The activity coefficient of every species is the sum of a combinatorial, a residual and a long-range part.
The residual part follows from the activity coefficients of the surface segments of all species, which interact pairwise in contact \citep{klamtConductorlikeScreeningModel1995,klamt_cosmospace_2002,gerlachOpenSourceCOSMORS2022}.
Two neutral molecules interact through contacts described by the simplified misfit and hydrogen bond described above.
Where an ion surface takes part, the misfit is scaled by a factor $A_k$, and the hydrogen bond is replaced by an empirical attractive energy scaled by a factor $B_k$ or is kept and scaled (Table~\ref{SI-tab:es-contacts} of the Supporting Information).
These energies distinguish monoatomic and polyatomic cations and anions, water and organic molecules \citep{gerlachDevelopmentCOSMORSBased2018,gonzalezdecastillaRevisionExtensionCOSMORS2025}.
Thresholds $C_0$ and $C_1$ shift the onset of two of them, and factors with $D$ and $E$ damp the attraction between anions and strongly charged cation segments.
The monoatomic ions are spheres with a uniform screening charge and fixed radii (Table~\ref{SI-tab:es-params} of the Supporting Information).

The long-range part is the ME-PDH term \citep{gonzalezdecastillaAnalogyRestrictedPrimitive2022,gonzalezdecastillaRevisionExtensionCOSMORS2025}.
It is a model based on the Pitzer--Debye--H\"uckel theory including the dependence of density, molar mass and permittivity on composition, and its closest-approach parameter grows at low permittivity to correct the screening there. The model also includes fundamental improvements in describing ions in mixtures, especially up to 100\% ionic liquid.
The distance of closest approach adds the radii of cation and anion, for a polyatomic ion the radius of the sphere with its COSMO volume times its sphericity, scaled by $F_\mathrm{M}$ for monoatomic and by $F_\mathrm{P}$ for polyatomic ions.
Density and permittivity mix the pure solvents and the ions by volume fraction, the ions with the density and permittivity of a hypothetical molten salt, so that the permittivity falls as salt is added.

COSMO-RS-ES differs from the published COSMO-RS-ES+ in its basis (Table~\ref{tab:es-vs-esplus}) and is therefore refitted (Section~\ref{sec:param} and Section~\ref{SI-sec:si_refit} of the Supporting Information).
It takes the $\sigma$-profiles from ORCA and the simplified short range of \citet{markgrafExtensionOpenCOSMORSFull2026}, whereas COSMO-RS-ES+ uses Turbomole profiles and the short range of \citet{gerlachDevelopmentCOSMORSBased2018}, in which the hydrogen bond depends on temperature and the misfit of neutral contacts carries a correction for the polarization of the surroundings.
Its combinatorial term keeps only the Flory--Huggins part, with the COSMO volumes.

\subsubsection{\ESPhi{}}\label{sec:methods-esphi}

\oPhi{} \citep{markgrafExtensionOpenCOSMORSFull2026}, the open-source form of COSMO-SAC-Phi \citep{dep.soaresPairwiseSurfaceContact2019}, treats a fluid as a pseudo-mixture of the real species and holes, which fill the free volume.
Every species $i$ has a cavity volume $b_i$, a hole volume $b_{h,i}$ and two dispersion parameters $\delta_i^0$ and $\delta_i^T$.
The volume is $V=\sum_in_ib_i+n_hb_h$ with the mean hole volume $b_h=\sum_ix_ib_{h,i}$, so that the amount of holes $n_h$ follows from $V$ and $\mathbf{n}$.
The residual Helmholtz energy combines a van der Waals repulsion with an attractive term from the openCOSMO-RS statistics of the pseudo-mixture, in which species and holes contribute surface segments and segment pairs interact through the simplified misfit, the hydrogen bond and a dispersion energy $-\tfrac12\sqrt{\delta_i\delta_j}$ with $\delta_i=\delta_i^0\left[1-\exp(-\delta_i^T/T)\right]$ \citep{dep.soaresPairwiseSurfaceContact2019,markgrafExtensionOpenCOSMORSFull2026}.

\ESPhi{} adds the ions as further species of this pseudo-mixture, each with its own cavity volume, hole volume and dispersion parameters, which interact with all other segments through the contact energies of COSMO-RS-ES, and adds the ME-PDH term as a Helmholtz energy,
\begin{equation}
A^\mathrm{r}(T,V,\mathbf{n})=
\underbrace{-nRT\ln\frac{V-\sum_in_ib_i}{V}}_{A^\mathrm{r}_\mathrm{R}}
+\underbrace{\sum_in_i\left(\tilde\mu^\mathrm{r}_{\mathrm{A},i}-\frac{b_i}{b_h}\tilde\mu^\mathrm{r}_{\mathrm{A},h}\right)-P_\mathrm{A}V}_{A^\mathrm{r}_\mathrm{A}}
\underbrace{-\,nRT\,\frac{4A_\phi I_x}{b_\mathrm{m}}\ln\left(1+b_\mathrm{m}I_x^{1/2}\right)}_{A^\mathrm{LR}},
\label{eq:helmholtz}
\end{equation}
where the sums run over all species including the ions, $\tilde\mu^\mathrm{r}_{\mathrm{A},i}$ and $\tilde\mu^\mathrm{r}_{\mathrm{A},h}$ are the attractive residual chemical potentials of species $i$ and of the holes in the pseudo-mixture, and $P_\mathrm{A}=-\tilde\mu^\mathrm{r}_{\mathrm{A},h}/b_h$ is the attractive pressure \citep{markgrafExtensionOpenCOSMORSFull2026}.
$A_\phi$, $b_\mathrm{m}$ and $I_x$ are those of Table~\ref{tab:es-equations}, evaluated at the molar density $n/V$ of the equation of state and with the permittivity of Eq.~\eqref{eq:eps}.
Pressure and fugacity coefficients follow from Eq.~\eqref{eq:helmholtz} by differentiation.
All calculations with \ESPhi{} are carried out as PTx flash calculations, in which the state is solved at the given pressure, temperature and composition.
Activity coefficients are ratios of fugacity coefficients at the same $T$ and $P$, with infinite dilution in the salt-free solvent of composition $x'$ as the reference state of the ions,
\begin{equation}
\ln\gamma_{\pm}^{(m)}=\frac{1}{\nu}\sum_i\nu_i\left[\ln\varphi_i(T,P,x)-\ln\varphi_i^\infty(T,P,x')\right]-\ln\left(1+\nu mM_\mathrm{s}\right),
\label{eq:lngamma}
\end{equation}
where $\nu_i$ is the number of ions $i$ per formula unit, $\nu=\sum_i\nu_i$, $m$ the molality and $M_\mathrm{s}$ the molar mass of the salt-free solvent.

The permittivity follows from volume and composition through the perturbation-theory model of \citet{ruebenPermittivityModelingElectrolyte2024},
\begin{equation}
\varepsilon_\mathrm{r}-1=3y\left[1+y+\left(\tfrac{17}{16}I-1\right)y^2\right],\quad
y=\frac{4\pi}{9}\sum_i\rho_i\left[\frac{(a_{11}\mu^2)_i}{k_\mathrm{B}T}+3(a_{12}\alpha)_i\right],\quad
I=1+\Big(\sum_ia_{2,i}x_i\Big)\left(e^{-y}-1\right),
\label{eq:eps}
\end{equation}
where $\rho_i$ is the number density of species $i$.
The coefficients of the solvents are taken from the database of \citet{ruebenPredictingRelativeStatic2024}.
The ion-specific coefficients of \citet{ruebenPermittivityModelingElectrolyte2024} are not used, because they were adjusted to MIACs within ePC-SAFT and therefore also compensate the errors of that equation of state.
Instead, every ion takes water's $a_2=0.1226$, which keeps the mixing of the correlation term neutral, and all ions share one polarizability $a_{12}\alpha$ adjusted to the dielectric decrement of the underlying mixing rule used in COSMO-RS-ES.

COSMO-RS-ES gives the ion energies no temperature dependence.
In \ESPhi{}, the 17 ion--solvent and the 9 ion--ion energies are therefore multiplied by
\begin{equation}
g(T)=\max\left\{0,\,1-w+w\,\frac{298.15~\mathrm{K}}{T}\right\},
\label{eq:gT}
\end{equation}
with one weight $w$ per group.
The factor has the form of the temperature factor of the hydrogen-bond energy in COSMO-RS \citep{klamt_cosmo-rs_2000} and leaves the model unchanged at 298.15~K.

\subsubsection{The Born term and dielectric decrement}\label{sec:methods-decrement-born}

Two electrolyte effects beyond the long-range term are examined with COSMO-RS-ES: the Born term, which describes the solvation of the ions, and the dielectric decrement, the change in permittivity when ions are added to the mixture.

The Born term goes back to the solvation energy of a charged sphere in a dielectric continuum \citep{bornVolumenUndHydratationswaerme1920}.
In electrolyte equations of state it carries the solvation of each ion through a cycle that discharges the ion, transfers it and charges it again \citep{bulowEPCSAFTAdvancedPart2021b}.
The form used here,
\begin{equation}
\ln\gamma_i^{\mathrm{Born}}=\frac{N_\mathrm{A}e^2}{8\pi\varepsilon_0RT}\left[\left(\frac{1}{\varepsilon_\mathrm{r}}-\frac{1}{\varepsilon_\mathrm{sf}}\right)\frac{z_i^2}{r_i}-\frac{1}{\varepsilon_\mathrm{r}^2}\left(\frac{\partial\varepsilon_\mathrm{r}}{\partial n_i}\right)_{T,n_{j\neq i}}\sum_{j\in\mathrm{ions}}\frac{n_jz_j^2}{r_j}\right],
\label{eq:born}
\end{equation}
describes only how the solvation of the ions changes as the salt changes the permittivity \citep{muellerApplicationRefinementCOSMORSES2020}.
Here $z_i$ is the charge number, $r_i$ the Born radius and $\varepsilon_\mathrm{sf}$ the permittivity of the salt-free solvent.
The Born radius sets the size of the term, which scales with $z_i^2/r_i$, and three sets are used.
COSMO radii, the COSMO-RS-ES radii of the monoatomic ions and for polyatomic ions the radius of a sphere with the area of their COSMO cavity, belong to the cavity on which COSMO-RS evaluates solvation and exist for every ion, so only they allow a fit on the full training data.
Crystal radii \citep{paulingSIZESIONSSTRUCTURE1927a} exist only for monoatomic ions and hydrated radii \citep{marcusThermodynamicsSolvationIons1991} not for the ions of ionic liquids, so fits with them are restricted to the salts of the ions these tables cover.

The dielectric decrement is the fall of the static permittivity of a high-permittivity solution with salt content, which is observed for e.g. aqueous salt solutions \citep{hastedDielectricPropertiesAqueous1948} and arises partly because water bound in the hydration shells of the ions no longer orients in a field \citep{buchnerDielectricRelaxationAqueous1999,maribo-mogensenModelingDielectricProperties2013a}.
In COSMO-RS-ES the decrement follows from the volume-fraction mixture of the salt-free solvent and the ionic melt (Table~\ref{tab:es-equations}), and without it the permittivity of the salt-free solvent is used at every point.
Without a decrement, $\varepsilon_\mathrm{r}=\varepsilon_\mathrm{sf}$ and the Born term of Eq.~\eqref{eq:born} is zero at every composition, so a Born term can only act together with a decrement.

\subsection{Experimental data}\label{sec:data}

The reach of an electrolyte model is set less by its equations than by the data available to parametrize and test it, and that constraint is strongly asymmetric. Mean ionic activity coefficients (MIACs) of aqueous alkali halide solutions at ambient temperature are abundant and mutually consistent, whereas mixed- and non-aqueous systems, temperature-resolved measurements and concentration-dependent solution properties are comparatively scarce, are frequently reported in incompatible conventions, and often disagree between sources. The present work therefore draws on several separately curated sources, each covering one of these regimes.
Table~\ref{tab:datasets} gives an overview of all data sets, their sources and their role.

\begin{table}[htb]
\centering
\caption{Data sets of this work.
Only the data marked as fitted enter the fit of \ESPhi{}; all others are predicted.}
\label{tab:datasets}
\small
\begin{tabular}{lp{0.30\textwidth}lll}
\hline
Data set & Source & $T$ & Size & Role \\
\hline
\multicolumn{5}{l}{\textbf{Aqueous solutions}} \\
MIAC, water                  & COSMO-RS-ES database \citep{gonzalezdecastillaRevisionExtensionCOSMORS2025} & 298.15~K      & 51 salts, 1419 points$^a$   & fitted \\
Density, water               & correlations of \citet{apelblatNewTwoparameterEquation2016} & 298.15~K      & 1419 points$^b$             & fitted \\
MIAC, water (JESS)           & JESS-FIZ \citep{mayJESSJointExpert2010} & 273--373~K    & 46 salts, 3535 points$^c$   & fitted \\
Permittivity, water          & measured static permittivities$^d$ & 298.15~K      & 24 salts, 188 points        & predicted \\
\addlinespace[1ex]
\multicolumn{5}{l}{\textbf{Mixed and organic solvents}} \\
MIAC, water + cosolvent      & COSMO-RS-ES database \citep{gonzalezdecastillaRevisionExtensionCOSMORS2025} & 298.15~K      & 88 systems, 1595 points$^a$ & predicted \\
MIAC, organic solvents       & COSMO-RS-ES database \citep{gonzalezdecastillaRevisionExtensionCOSMORS2025} & 298.15~K      & 22 systems, 508 points$^{a,e}$  & predicted \\
MIAC, Yang database          & \citet{yangBenchmarkDatabaseMixedSolvent2022} & 288--318~K    & 120 series, 1986 points     & predicted \\
Bubble pressure, Yang        & \citet{yangBenchmarkDatabaseMixedSolvent2022} & isothermal    & 9 systems, 157 points       & predicted \\
Bubble temperature, Yang     & \citet{yangBenchmarkDatabaseMixedSolvent2022} & isobaric      & 11 systems, 217 points      & predicted \\
\addlinespace[1ex]
\multicolumn{5}{l}{\textbf{Phase equilibria with salts}} \\
Salt solubility              & electrolyte database of COSMO-RS-ES & 298.15~K      & 125 systems, 364 points     & predicted \\
LLE with salts               & training data of COSMO-RS-ES & 298.15~K      & 93 systems, 701 tie lines   & predicted$^f$ \\
\hline
\multicolumn{5}{p{0.95\textwidth}}{\footnotesize $^a$ The systems of the training data of COSMO-RS-ES (Table~\ref{tab:es-vs-esplus}) whose salt has a density correlation, in water, methanol, ethanol, 1-propanol, 2-propanol, acetonitrile, ethylene glycol, 1,4-dioxane and 1,2-dimethoxyethane, up to 20~mol~kg$^{-1}$.} \\
\multicolumn{5}{l}{\footnotesize $^b$ 1371 of them inside the range of the density correlations.} \\
\multicolumn{5}{l}{\footnotesize $^c$ Up to 6~mol~kg$^{-1}$, 2376 of them in the fit (Section~\ref{sec:param}).} \\
\multicolumn{5}{p{0.95\textwidth}}{\footnotesize $^d$ \citet{hastedDielectricPropertiesAqueous1948,haggisDielectricPropertiesWater1952,harrisDielectricPropertiesAqueous1957,hastedDielectricPropertiesAqueous1958,barthelUntersuchungenDispersionKomplexen1970,barthelUntersuchungenDispersionKomplexen1977}, corrected for kinetic depolarization \citep{hubbardDielectricDispersionDielectric1977}.} \\
\multicolumn{5}{l}{\footnotesize $^e$ 37 systems and 901 points with 14 systems flagged as untrusted.} \\
\multicolumn{5}{l}{\footnotesize $^f$ Fitted by COSMO-RS-ES.} \\
\hline
\end{tabular}
\end{table}

\paragraph{Aqueous solutions}
The aqueous and mixed-aqueous MIAC data are taken from the database assembled for the COSMO-RS-ES model \citep{gerlachDevelopmentCOSMORSBased2018,gonzalezdecastillaAnalogyRestrictedPrimitive2022,gonzalezdecastillaRevisionExtensionCOSMORS2025}, whose most recent revision comprises 3657 data points in 184 systems at 298.15~K, of which roughly 60~\% involve alkali halides. About 40~\% of these points refer to pure water and a further 44~\% to mixed aqueous solvents, so the ion-specific short-range parameters are effectively determined by aqueous and alcohol-rich aqueous systems.
The densities of the aqueous solutions are not measured points but values of the correlations of \citet{apelblatNewTwoparameterEquation2016}, evaluated at the composition of every MIAC point, and salts without such a correlation are left out of the fit.
The temperature dependence is evaluated against MIAC of 46 salts in water from the thermo-physical database JESS-FIZ of the Joint Expert Speciation System \citep{mayJESSJointExpert2010}, which \citet{gonzalezdecastillaRevisionExtensionCOSMORS2025} also used to test COSMO-RS-ES+, and of which the data between 273 and 373~K up to 6~mol~kg$^{-1}$ are used.
The permittivities of aqueous salt solutions are measured static permittivities with the kinetic depolarization removed by the model of \citet{hubbardDielectricDispersionDielectric1977} (Section~\ref{sec:permittivity}).

\paragraph{Water--alcohol mixtures}
For the evaluation in water--alcohol mixtures, the benchmark database of \citet{yangBenchmarkDatabaseMixedSolvent2022} is used. It covers (water + methanol/ethanol + alkali halide) mixtures and reconciles MIAC and vapor-liquid equilibrium (VLE) data against one another rather than evaluating them in isolation. Starting from 61 data sets collected from DETHERM, the CERE electrolyte databank and other literature sources, the authors analyze the data with an E-NRTL model, remove inconsistent sets and classify the remaining ones as recommended, tentative or uncertain. The recommended and tentative sets form the benchmark, 1413 data records from 32 data sets for 13 solutions, and the uncertain sets, supplied separately, are evaluated here as well. Since MIAC and VLE constrain the composition derivatives of the excess Gibbs energy with respect to the salt and to the solvent respectively, the two properties together test a mixed-solvent model considerably more stringently than either alone, which is why both are used here.

\paragraph{Phase equilibria with salts}
Salt solubilities and liquid--liquid equilibria with salts are evaluated as in an earlier publication of COSMO-RS-ES on salt solubilities in solvent mixtures \citep{mullerEvaluationRefinementNovel2019a}. Both are calculated as described by \citet{gonzalezdecastillaAnalogyRestrictedPrimitive2022}.
The solubilities of single salts of the 20 ions of \ESPhi{} in water, water + cosolvent and pure organic solvents with \oPhi{} parameters come from the electrolyte database of the COSMO-RS-ES framework.
The tie lines between an aqueous and an organic-rich phase are the LLE data with salts on which COSMO-RS-ES is fitted, restricted to the same ions and to solvents with published parameters \citep{markgrafExtensionOpenCOSMORSFull2026}.

\paragraph{Database}
The experimental data of this work are part of an electrolyte database that is published openly together with this article \citep{mullerElectrolyteDatabase2026}.
Beyond the systems used here, it compiles MIAC, activity coefficients at infinite dilution, SLE, LLE and VLE of electrolyte systems including ionic liquids, Gibbs energies of transfer of ions and permittivities of electrolyte solutions.
The structures of the salts were checked with MoleculeResolver \citep{mullerHowCrackSMILES2025}, which resolves names and CAS numbers to structures and cross-checks them across several chemical databases.

\subsection{Parametrization}\label{sec:param}

The parametrization builds up the electrolyte equation of state in three steps.
First, the neutral model is taken from \citet{markgrafExtensionOpenCOSMORSFull2026} unchanged, with its global parameters and the pure-component parameters of all solvents.
Second, COSMO-RS-ES is refitted on this neutral basis, since the parameters of COSMO-RS-ES+ \citep{gonzalezdecastillaRevisionExtensionCOSMORS2025} belong to a different short range model.
Third, the refitted COSMO-RS-ES is the basis for the extension into \ESPhi{}, which takes over its ion interaction energies and adjusts only the volume, dispersion and temperature parameters of the ions.

\subsubsection{Objective function}\label{sec:objective}

\begin{description}[style=nextline]
\item[COSMO-RS-ES]
COSMO-RS-ES is refitted on nearly the same training data as \citet{gonzalezdecastillaRevisionExtensionCOSMORS2025}, which comprise MIAC in water, water + cosolvent and pure organic solvents, LLE with salts and ionic liquids, activity coefficients at infinite dilution in ionic liquids and Gibbs energies of transfer, and with the same form of the objective function.
The weights per data group differ from those of \citet{gonzalezdecastillaRevisionExtensionCOSMORS2025} and are 300, 150 and 5 for the MIAC in water, water + cosolvent and pure organic solvents, 10 and 2.5 for the LLE with salts and with ionic liquids, 0.65 for the activity coefficients at infinite dilution and 0.1 for the Gibbs energies of transfer (Table~\ref{tab:es-vs-esplus}).

\item[\ESPhi{}]
\ESPhi{} is fitted on aqueous data only, the MIAC at 298.15~K, their densities and the JESS data.
Its objective sums the squared deviations of $\ln\gamma_\pm$ and of the density over all training points $k$,
\begin{equation}
\mathrm{OF}=\sum_k\big(\ln\gamma_{\pm,k}^{\mathrm{calc}}-\ln\gamma_{\pm,k}^{\mathrm{exp}}\big)^2+w_\rho\sum_k\big(\rho_k^{\mathrm{calc}}-\rho_k^{\mathrm{corr}}\big)^2,
\label{eq:of}
\end{equation}
with the MIAC on the molality scale of Eq.~\eqref{eq:lngamma}, the densities $\rho_k^{\mathrm{corr}}$ of the correlations of \citet{apelblatNewTwoparameterEquation2016} in g~cm$^{-3}$ and the density weight $w_\rho=15$.
Where the 298.15~K set contains a salt, its JESS isotherm at 298.15~K is left out, so that no salt enters twice.
The objective is minimized with the Nelder--Mead method \citep{nelder_simplex_1965} from multiple starting points, and the best result is the final parameter set.
\end{description}

\section{Results and discussion}\label{sec:results}

The results follow the three steps of the parametrization (Section~\ref{sec:param}).
The neutral species keep the parameters of \oPhi{} \citep{markgrafExtensionOpenCOSMORSFull2026} throughout, so that no solvent or solvent mixture is refitted.
On this basis, the electrolyte activity coefficient model COSMO-RS-ES is first refitted to the simplified openCOSMO-RS, and with it the Born term and the dielectric decrement are tested (Section~\ref{sec:results-es}).
The refitted COSMO-RS-ES supplies the ion interaction energies of \ESPhi{}. The only parameters of the equation of state that are fitted for the ions in the last step are the pure-component parameters: the cavity volume, the corresponding hole volume and the dispersion parameters (Section~\ref{sec:results-esphi}).

\subsection{COSMO-RS-ES}\label{sec:results-es}

For COSMO-RS-ES a total of 33 general parameters are adjusted. Their values are listed in Table~\ref{SI-tab:es-params} of the Supporting Information.
Table~\ref{tab:es-vs-esplus} compares the refitted model with the published COSMO-RS-ES+.
On the training data the current COSMO-RS-ES deviates more than COSMO-RS-ES+ on the different data types except the MIAC in pure organic solvents and the Gibbs energies of transfer.
The cause of this difference is most probably the simplified short range equation used, which already makes the neutral model slightly less accurate (Section~\ref{sec:models}).

\begin{table}[!htb]
\centering
\caption{Basis, fit and AAD on the training data of COSMO-RS-ES (this work) and COSMO-RS-ES+ \citep{gonzalezdecastillaRevisionExtensionCOSMORS2025}, with the number of points in parentheses.
The AAD refers to $\ln\gamma_\pm$ for MIAC, $\ln K$ for LLE, $\ln\gamma^\infty$ for IDAC and $\Delta G/RT$ per ion for transfer Gibbs energies.
Units: $\alpha$ and $c_\mathrm{hb}$ in kJ~\AA$^2$~mol$^{-1}$~$e^{-2}$, $\sigma_\mathrm{hb}$ in $e$~\AA$^{-2}$.}
\label{tab:es-vs-esplus}
\small
\begin{tabular}{p{0.34\textwidth}p{0.28\textwidth}p{0.28\textwidth}}
\hline
 & This work & COSMO-RS-ES+ \\
\hline
$\sigma$-profiles & ORCA 6.0, TZVPD & Turbomole 6.6, BP86/def-TZVP \\
Short range of neutral species & simplified openCOSMO-RS \citep{markgrafExtensionOpenCOSMORSFull2026} & full COSMO-RS \citep{gerlachDevelopmentCOSMORSBased2018} \\
$a_\mathrm{eff}$, $\alpha$, $c_\mathrm{hb}$, $\sigma_\mathrm{hb}$ & 4.857~\AA$^2$, 8130, 42\,564, 0.00944 & 6.25~\AA$^2$, 5950, 36\,700, 0.0085 \\
Temperature factor $c_\mathrm{hb}^T$ & 0 & 1.5 \\
Misfit of neutral contacts & without correlation term & with correlation term, $f^\perp=2.4$ \\
Hydrogen atoms distinguished by bonding partner & no & yes \\
Combinatorial term & Flory--Huggins & Staverman--Guggenheim, $z=10$ \\
Radii of monoatomic ions & fixed & fitted \\
Weights, MIAC in water, water + cosolvent and organic solvents & 300, 150, 5 & 60, 40, 15 \\
Weights, LLE with salts and ionic liquids & 10, 2.5 & 3, 2 \\
Weights, IDAC and transfer energies & 0.65, 0.1 & 0.7, 0.3 \\
\hline
AAD, MIAC in water & 0.103 (1479) & 0.063 (1479) \\
AAD, MIAC in water + cosolvent & 0.061 (1595) & 0.050 (1595) \\
AAD, MIAC in organic solvents & 0.331 (583) & 0.416 (583) \\
AAD, LLE with salts & 0.827 (1143) & 0.649 (1056) \\
AAD, LLE with ionic liquids & 1.169 (1481) & 1.025 (1637) \\
AAD, IDAC in ionic liquids & 0.828 (21\,652) & 0.655 (21\,634) \\
AAD, Gibbs energies of transfer & 2.956 (1189) & 3.052 (1147) \\
\hline
\end{tabular}
\end{table}

\subsubsection{The Born term and dielectric decrement}\label{sec:decrement-born}

The Born term and the dielectric decrement are tested with the activity coefficient model rather than with the equation of state.
In the development of \oPhi{}, the equation of state behaved essentially like the underlying activity coefficient model: wherever the activity coefficient model predicts an azeotrope, the equation of state does as well, and wherever it does not, neither does the equation of state \citep{markgrafExtensionOpenCOSMORSFull2026}.
The same was observed for the electrolyte extension.
Since COSMO-RS-ES converges much faster than \ESPhi{}, the tests of a Born term and of a dielectric decrement are carried out with COSMO-RS-ES, and what is learned from them applies to \ESPhi{} as well.
To this end, COSMO-RS-ES is fitted in three configurations.
The first uses the decrement without the Born term and gives the parameter set of COSMO-RS-ES used throughout this work, the second omits the decrement, so that the Born term vanishes as well, and the third uses both.

COSMO-RS starts from the COSMO screening charges, which already carry the dielectric screening of a solute \citep{klamtCOSMONewApproach1993}, and replaces the dielectric continuum by the statistical thermodynamics of interacting surface segments \citep{klamtCOSMOCOSMORSSolvation2018}.
COSMO-RS-ES describes the solvation of the ions within this framework \citep{gerlachDevelopmentCOSMORSBased2018}, so a Born term is expected to count it a second time.
\citet{mullerEvaluationRefinementNovel2019a} already asked for this reason whether a Born term should be used and left it out, and in an earlier study within COSMO-RS-ES, evaluated on aqueous MIAC, a decrement in the long-range term improved the model only marginally, whereas every Born variant made it clearly worse \citep{muellerApplicationRefinementCOSMORSES2020}.
Since the decrement lowers $\varepsilon_\mathrm{r}$, the Born term raises $\ln\gamma_\pm$ with salt, most strongly where phases differ strongly in permittivity.
As Born's model matches hydration energies only with radii enlarged beyond the crystal radii \citep{latimerFreeEnergyHydration1939,rashinReevaluationBornModel1985}, the hydrated radii, larger than the crystal radii and therefore giving a weaker term, are expected to do best.
For the dielectric decrement, a lower permittivity strengthens the interaction between the ions in the long-range term and makes $\ln\gamma_\pm$ more negative at high molality, which shifts the balance between the long-range and the short-range term.
A decrement is therefore expected to help mainly in concentrated aqueous solutions.
Its effective size in a thermodynamic model is expected to stay below the measured one, because measured values also contain kinetic depolarization \citep{maribo-mogensenModelingDielectricProperties2013a} and because in earlier fits to thermodynamic data together with a flexible short-range term the decrement came out weaker than measured \citep{muellerApplicationRefinementCOSMORSES2020,olsenComparisonsEquationState2023}.

\begin{figure}[htb]
\centering
\includegraphics[width=\textwidth]{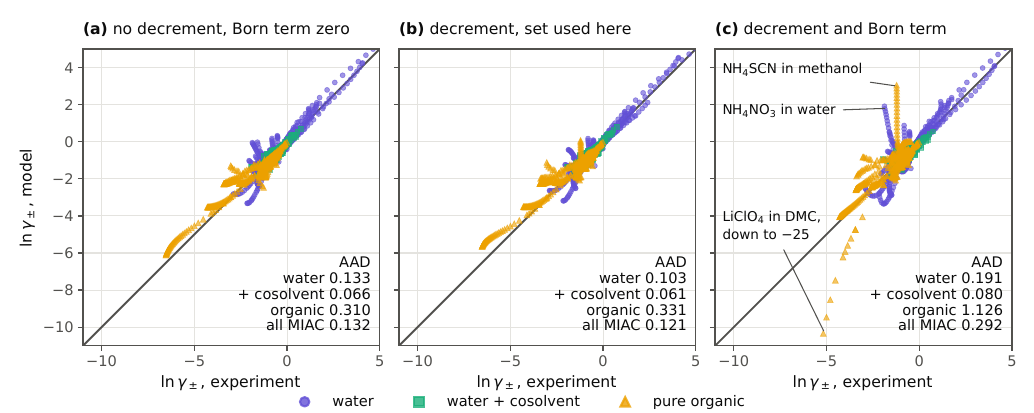}
\caption{MIAC of the COSMO-RS-ES training data, model against experiment, fitted (a) without dielectric decrement, and hence without Born term, (b) with decrement, as used in this work, and (c) with decrement and Born term (COSMO radii).
Circles: water; squares: water + cosolvent; triangles: pure organic solvents.
In (c), LiClO$_4$ in DMC continues below the axis to $-25$.}
\label{fig:decrement-born-parity}
\end{figure}

Figure~\ref{fig:decrement-born-parity} compares the MIAC of the training data for the best fit of each configuration.
Without the decrement the Born term of Eq.~\eqref{eq:born} is zero, so the fit without the decrement is also a fit without a Born term, and the Born term can only be tested on top of the decrement.
The decrement improves the aqueous MIAC, with an AAD of 0.103 against 0.133 without it, and those in water + cosolvent, 0.061 against 0.066.
Only the MIAC in pure organic solvents, weighted 5 against 300 for water, do not improve (0.331 against 0.310).
Over all MIAC the AAD falls from 0.132 to 0.121, and the other data groups of the training set improve as well, most the LLE with ionic liquids.
Every completed fit with the decrement ends at a lower objective than every fit without it, so the ranking does not depend on the choice of the best fits.
For NaCl at 6~mol~kg$^{-1}$, the decrement lowers the permittivity from 78.5 to 66.7 and makes the long-range part of $\ln\gamma_\pm$ 0.40 more negative, the refitted short range compensates, and the sum, $-0.135$ against $-0.280$, lies closer to the experimental $-0.014$.
Without the decrement, the aqueous loss comes mostly from the sulfates, Cs$_2$SO$_4$, Rb$_2$SO$_4$ and (NH$_4$)$_2$SO$_4$, and from Li and Na salts that reach high molality, LiBr and LiNO$_3$ up to 20~mol~kg$^{-1}$ and NaI up to 12~mol~kg$^{-1}$, whereas LiCl, KSCN and CsCl are fitted better without it.
The streaks that appear in all three panels, such as LiNO$_3$ and LiClO$_4$ in 2-propanol and (NH$_4$)$_2$SO$_4$ in water, are errors of COSMO-RS-ES itself.

The Born term makes every data group worse, the aqueous MIAC from 0.103 to 0.191, those in water + cosolvent from 0.061 to 0.080 and those in pure organic solvents from 0.331 to 1.126 (Figure~\ref{fig:decrement-born-parity}c).
Three properties make the term large.
The first is a low permittivity, because the term scales with $1/\varepsilon_\mathrm{r}-1/\varepsilon_\mathrm{sf}$.
In dimethyl carbonate (DMC), with a salt-free permittivity of 3.1, the ionic melt even raises the permittivity, so the term turns negative and LiClO$_4$ in DMC falls to $-25$, whereas the data go down only to $-6.5$.
This single system carries 84~\% of the loss in pure organic solvents.
The second is a high molality, at which the decrement and with it the term are largest, as for NH$_4$NO$_3$ up to 20~mol~kg$^{-1}$ in water.
The third is a double charge, which quadruples $z_i^2$, as for the sulfates of Cs and Rb and the hydrogen phosphates.
Where the term is large, the refitted parameters move to absorb it and thereby spoil other systems, such as NH$_4$SCN in methanol, whose model value rises to $+3$ although its own Born term stays below 0.6.
The largest deterioration lies outside the MIAC, in the LLE with ionic liquids, where the term of free ions reaches tens of logarithmic units between an IL-rich and an apolar phase of very different permittivity, with medians of 12.6 and 3.1, in line with the strong sensitivity of the LLE of ionic liquids to the description of the long-range electrostatics \citep{marquesImpactExtendedLongrange2023}.
A plausible reason, not tested here, is that ionic liquids dissolve in apolar solvents as ion pairs, which a free-ion term cannot describe.

On the 104 alkali-halide data sets in all solvents (2190 points), which all three radius sets cover, the model without the Born term reaches an AAD of 0.107, the hydrated radii 0.097, the crystal radii 0.113 and the COSMO radii 0.114.
As expected the hydrated radii do best, but their fit had fewer data groups to satisfy, and the only like-for-like comparison, COSMO radii against no Born term on the same data, is worse even here.
A larger radius thus only weakens a term that counts the solvation twice, and no radius is expected to improve the model as a whole.
Thus, the Born term is neglected in both models of this work, while the dielectric decrement is used, as in COSMO-RS-ES+ \citep{gonzalezdecastillaRevisionExtensionCOSMORS2025}.

The improvement does not prove that the correlated decrement is physical, because the refitted short range absorbs much of the change \citep{kontogeorgisDebyeHuckelTheoryIts2018}. However, it means that describing this effect, albeit qualitatively, improves the predictive power of the model.
Against measured static permittivities of aqueous salts, corrected for kinetic depolarization \citep{hubbardDielectricDispersionDielectric1977}, COSMO-RS-ES lies above the data for the Li and Na salts (Figure~\ref{fig:eps}) and reproduces only about a third of the corrected decrement of NaCl at 1~mol~kg$^{-1}$. A permittivity model with dielectric saturation, such as the model of \citet{maribo-mogensenModelingDielectricProperties2013a}, would replace the effective decrement, and a Born-type term on the permittivity of the salt-free solvent, which vanishes in water, would act exactly where transfer energies between solvents are defined.

\subsection{\ESPhi{}}\label{sec:results-esphi}

\subsubsection{Adjusted parameters}\label{sec:adjusted}

\begin{figure}[htb]
\centering
\includegraphics[width=\textwidth]{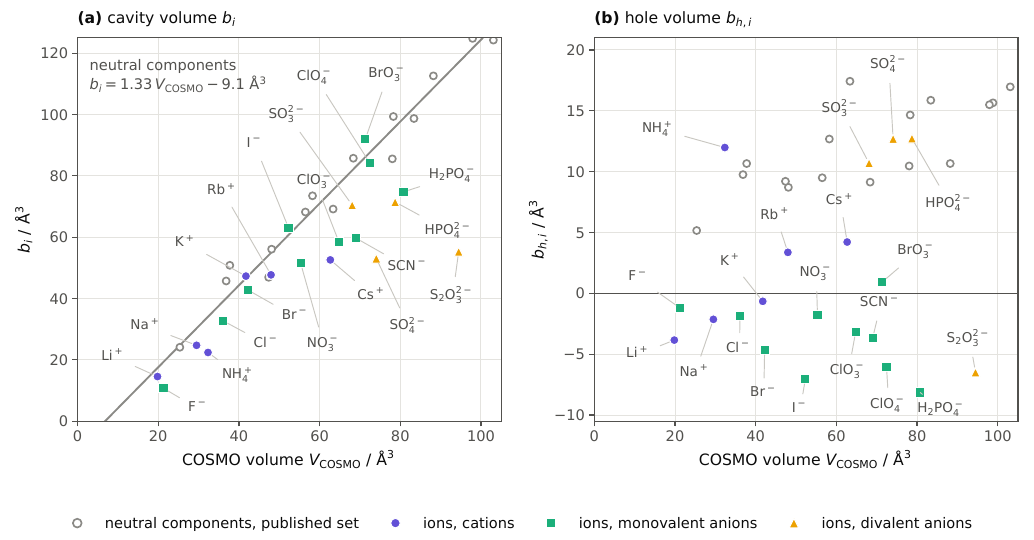}
\caption{Fitted cavity volumes (a) and hole volumes (b) of the ions against their COSMO volume, compared with the neutral components of \oPhi{} \citep{markgrafExtensionOpenCOSMORSFull2026}.
The gray line is the linear fit of the cavity volume of the neutral components.}
\label{fig:ion-params}
\end{figure}

In \ESPhi{}, the ion interaction energies, the closest approach, the permittivity coefficients and all solvent parameters are kept at the values from COSMO-RS-ES.
For every ion, the cavity volume $b_i$ and the hole volume $b_{h,i}$ are adjusted.
The dispersion is described by one $\delta^0$ shared by all cations and one shared by all anions, and every ion takes water's $\delta^T$ of 508.7~K.
Three considerations motivate this.
First, $\delta^0$ and $\delta^T$ are degenerate at a single temperature, so isothermal data can identify at most one combination of them.
Second, fitted per ion, $\delta^0$ is not determined by the data, since equally good fits from different starts scatter more than the values differ between the ions, the values run opposite to the polarizabilities of the ions, and the gain on the training data does not carry over to the temperature data.
Third, the attraction a per-ion value would describe is absorbed into the ion-specific interaction energies, which were fitted to the same kind of data, so adjusting it would mainly absorb further error without a gain in predictivity.
Cations and anions are kept apart as the simplest grouping beyond a single common value.
With the two weights $w$ of Eq.~\eqref{eq:gT}, this gives 44 adjusted parameters, whose fitted values are listed in Table~\ref{tab:params} and shown in Figure~\ref{fig:ion-params}.

\begin{table}[htb]
\centering
\caption{Fitted cavity and hole volumes of the 20 ions, with their COSMO volume (for the alkali cations, a sphere of their fixed radius) and the number of aqueous training salts containing each ion; volumes in \AA$^3$.
All cations share $\delta^0=1.707$~kcal~mol$^{-1}$, all anions $\delta^0=0.0007$~kcal~mol$^{-1}$, and all ions take the $\delta^T$ of water, 508.7~K.
The weights of Eq.~\eqref{eq:gT} are $-0.900$ for the ion--solvent and $-0.310$ for the ion--ion energies.
Values marked $^*$ lie at or within 0.1 of a bound of the fit, as does the $\delta^0$ of the anions.}
\label{tab:params}
\small
\begin{tabular}{lrrrr@{\hspace{2.5em}}lrrrr}
\hline
Cation & $V_\mathrm{COSMO}$ & $b_i$ & $b_{h,i}$ & Salts & Anion & $V_\mathrm{COSMO}$ & $b_i$ & $b_{h,i}$ & Salts \\
\hline
Li$^+$     & 19.8 & 14.56 & $-3.84$     &  7 & F$^-$                 & 21.3 & 10.74$^*$ & $-1.22$ & 4 \\
Na$^+$     & 29.5 & 24.74 & $-2.13$     & 14 & Cl$^-$                & 36.1 & 32.73 & $-1.83$ & 6 \\
K$^+$      & 41.7 & 47.30 & $-0.65$     & 12 & Br$^-$                & 42.2 & 42.80 & $-4.63$ & 6 \\
Rb$^+$     & 48.0 & 47.71 & 3.37        &  6 & I$^-$                 & 52.3 & 63.03 & $-7.02$ & 5 \\
Cs$^+$     & 62.7 & 52.59 & 4.22        &  6 & NO$_3^-$              & 55.3 & 51.47 & $-1.76$ & 6 \\
NH$_4^+$   & 32.3 & 22.40 & 11.99$^*$   &  6 & ClO$_3^-$             & 64.8 & 58.49 & $-3.16$ & 3 \\
           &      &       &             &    & ClO$_4^-$             & 72.4 & 84.24 & $-6.05$ & 4 \\
           &      &       &             &    & BrO$_3^-$             & 71.4 & 92.08 & 0.93    & 2 \\
           &      &       &             &    & SCN$^-$               & 69.0 & 59.71 & $-3.64$ & 2 \\
           &      &       &             &    & H$_2$PO$_4^-$         & 80.8 & 74.75 & $-8.15^*$ & 2 \\
           &      &       &             &    & HPO$_4^{2-}$          & 78.7 & 71.48 & 12.73$^*$ & 3 \\
           &      &       &             &    & SO$_3^{2-}$           & 68.1 & 70.44 & 10.70   & 1 \\
           &      &       &             &    & SO$_4^{2-}$           & 74.1 & 53.06 & 12.70$^*$ & 6 \\
           &      &       &             &    & S$_2$O$_3^{2-}$       & 94.5 & 55.29 & $-6.48^*$ & 1 \\
\hline
\end{tabular}
\end{table}

For most neutral components the cavity volume grows linearly with the COSMO volume \citep{dep.soaresPairwiseSurfaceContact2019,markgrafExtensionOpenCOSMORSFull2026}, here as $b_i=1.33\,V_\mathrm{COSMO}-9.1$~\AA$^3$, and the hole volume stays positive, between 5.2~\AA$^3$ for water and 19~\AA$^3$.
The five alkali cations, whose COSMO volumes rise with their fixed radii, lie close to this line (Table~\ref{SI-tab:es-params} of the Supporting Information).
Most other ions, however, depart from it toward smaller volumes.
Twelve of the 14 anion cavities lie below the line, most clearly for the small and the doubly charged anions, and 13 of the 20 hole volumes are negative.
This reflects electrostriction: the field of an ion orients and compresses the surrounding water, so that a dissolved ion takes up less volume than a neutral molecule of the same size, and the standard partial molar volume of Na$^+$ in water is even negative \citep{milleroMolalVolumesElectrolytes1971a,marcusElectrostrictionElectrolyteSolutions2011a,marcusStandardPartialMolar2004}.
\ESPhi{} can describe this contraction in two ways, by a smaller cavity or by a negative hole volume, which lowers the mean hole size $b_h=\sum_ix_ib_{h,i}$ of the mixture.
How the fit divides the contraction between the two is not unique, so the individual values are effective.
Some trends nevertheless follow the hydration of the ions.
The hole volume of the cations rises from Li$^+$ to Cs$^+$ as their hydration weakens \citep{zavitsasPropertiesWaterSolutions2001,marcusThermodynamicsSolvationIons1991}, and the halide cavities grow from F$^-$, at its lower bound of half its COSMO volume, to I$^-$ on the neutral line, although the halide hole volumes run the other way.

The negative hole volumes are essential for concentrated solutions, because without them the free volume cannot shrink as salt is added.
A refit with all hole volumes held non-negative raises the aqueous AAD from 0.073 to 0.098, and above 12~mol~kg$^{-1}$ from 0.14 to 0.44.
A few ions instead receive large positive hole volumes.
A salt whose ions have hole volumes below that of water, 5.17~\AA$^3$, shrinks the free volume and raises $\ln\gamma_\pm$ at high molality, and a partner ion with a large positive hole volume counteracts this.
The fit raises the hole volume of NH$_4^+$ until NH$_4$Cl and NH$_4$NO$_3$ leave the mean hole size of water practically unchanged, and pushes SO$_4^{2-}$ and HPO$_4^{2-}$ to the upper bound, where their lithium and sodium salts are still predicted too high at the highest molalities.
These values therefore show how far the fit had to push the parameters rather than the size of the ions, and the hole volume of SO$_3^{2-}$, fitted to a single salt, is almost undetermined.

The classification of ions as structure makers and breakers \citep{nguyenCorrelationsDensitiesAqueous2016a,reynoldsStructureMakerstructureBreaker2024}, available for 15 of the 20 ions, explains the fitted volumes only in part.
Structure makers bind water more strongly than water binds to itself, and the looser water around a breaker gives it a larger apparent volume.
The hole volumes of the alkali cations rise in this order, from the maker Li$^+$ to the breakers K$^+$, Rb$^+$ and Cs$^+$, but the breakers are also the larger ions, so the two effects cannot be separated.
Among the halides only the cavities follow the order, from the borderline F$^-$ to the breaker I$^-$, and the other monovalent anions follow neither volume consistently.
The large positive hole volumes of SO$_4^{2-}$ and HPO$_4^{2-}$ do not mark structure breakers either, since HPO$_4^{2-}$ is a maker and SO$_4^{2-}$ borderline, whereas the breaker S$_2$O$_3^{2-}$ has a negative hole volume.
These values follow from the free-volume balance above rather than from the water structure.

The dispersion parameter shared by the cations, 1.707~kcal~mol$^{-1}$, lies within the range of the neutral components, 0.40 to 2.51~kcal~mol$^{-1}$, whereas that of the anions runs to its lower bound of zero, so the attraction of the anions is carried by their interaction energies alone.
As for the neutral components, whose dispersion parameters show no clear correlation with size \citep{markgrafExtensionOpenCOSMORSFull2026}, these are effective parameters.
A segment-based dispersion term derived from atomic polarizabilities \citep{grigorashComprehensiveApproachIncorporating2025,grigorashComprehensiveApproachIncorporating2026} could tie them to a physical descriptor.

\subsubsection{Permittivity}\label{sec:permittivity}

The ion coefficients of the permittivity model are not fitted to thermodynamic data but to the permittivity of COSMO-RS-ES, the volume-fraction mixture of the salt-free solvent and the ionic melt (Table~\ref{tab:es-equations}).
\ESPhi{} thus follows the decrement with which the interaction energies it takes over were fitted, which is needed, since a different decrement describes the aqueous MIAC clearly worse (see below).
With every ion at the $a_2$ of water and one shared polarizability in Eq.~\eqref{eq:eps}, the fit over the 1419 points of the 51 aqueous training salts gives $a_{12}\alpha=+3.38$~\AA$^3$, within the range of real ion polarizabilities.
The dissolved salt then acts as a polarisable component without permanent dipoles, with an apparent permittivity of 6 to 10, close to the 5.45 of the ionic melt of COSMO-RS-ES.
Simpler choices match COSMO-RS-ES worse: ions without any coefficient give too weak a decrement, and ions at $a_2=0$ need a negative polarizability of $-8.14$~\AA$^3$.
Only a correlation parameter fitted per ion matches better, at the cost of one parameter per ion.

\begin{figure}[!htb]
\centering
\includegraphics[width=\textwidth]{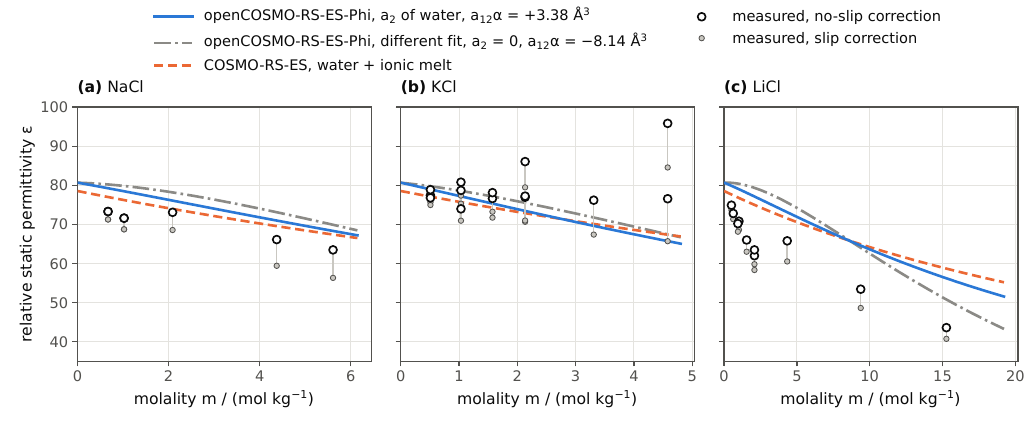}
\caption{Static permittivity at 298.15~K of (a) NaCl, (b) KCl and (c) LiCl.
Lines: \ESPhi{} (blue, solid), \ESPhi{} with the ions at $a_2=0$ (gray, dash-dotted) and COSMO-RS-ES (orange, dashed).
Circles: measured values \citep{hastedDielectricPropertiesAqueous1948,haggisDielectricPropertiesWater1952,harrisDielectricPropertiesAqueous1957,hastedDielectricPropertiesAqueous1958,barthelUntersuchungenDispersionKomplexen1970}, corrected for kinetic depolarization \citep{hubbardDielectricDispersionDielectric1977} with no-slip (open) and slip (small, gray) boundary conditions, using $\tau=8.7$~ps and $\varepsilon_\infty=5$.}
\label{fig:eps}
\end{figure}

Figure~\ref{fig:eps} compares the resulting permittivity with COSMO-RS-ES and with measured values for NaCl, KCl and LiCl, and Figure~\ref{SI-fig:eps-six} of the Supporting Information adds LiBr, NaSCN and NH$_4$NO$_3$.
The kinetic depolarization of the measured values is removed with the continuum model of \citet{hubbardDielectricDispersionDielectric1977} for both no-slip and slip boundary conditions, since dielectric spectra of NaCl point to slip \citep{buchnerDielectricRelaxationAqueous1999}, whereas simulations find the model valid only below about 0.2~mol~L$^{-1}$ \citep{segaKineticDielectricDecrement2015}.
Both models lie above the corrected data for NaCl, LiCl and LiBr, \ESPhi{} by 5.5 to 12.8 and COSMO-RS-ES by 3.6 to 9.0 with the no-slip correction and by somewhat more with the slip correction, while for KCl they lie about 5 below the no-slip values and match the slip values.
The data themselves are uncertain, as the KCl sources scatter by several units and LiCl was measured above 2.1~mol~kg$^{-1}$ by a single source, so they constrain the decrement of the models only loosely.

The curves of \ESPhi{} start at 80.7 instead of 78.4, because the model water is 2.3~\% too dense \citep{wagnerIAPWSFormulation1995Thermodynamic2002}, and then fall about linearly with molality, like those of COSMO-RS-ES.
Only for NaSCN and NH$_4$NO$_3$ at 18 to 20~mol~kg$^{-1}$ do they fall about 10 lower (Figure~\ref{SI-fig:eps-six}), because \ESPhi{} gives these salts 10 to 15~\% more volume than the molten salt of COSMO-RS-ES.
The fit with the ions at $a_2=0$ gives a decrement that starts flat and steepens, and it describes the aqueous MIAC worse, with an AAD of 0.086 instead of 0.073 and about twice the error above 7~mol~kg$^{-1}$.
Like that of COSMO-RS-ES, the permittivity of \ESPhi{} is therefore an effective quantity (Section~\ref{sec:decrement-born}).

\subsubsection{Aqueous solutions}\label{sec:aqueous}

\paragraph{Mean ionic activity coefficients}\label{sec:aqueous-miac}

\begin{figure}[htb]
\centering
\includegraphics[width=\textwidth]{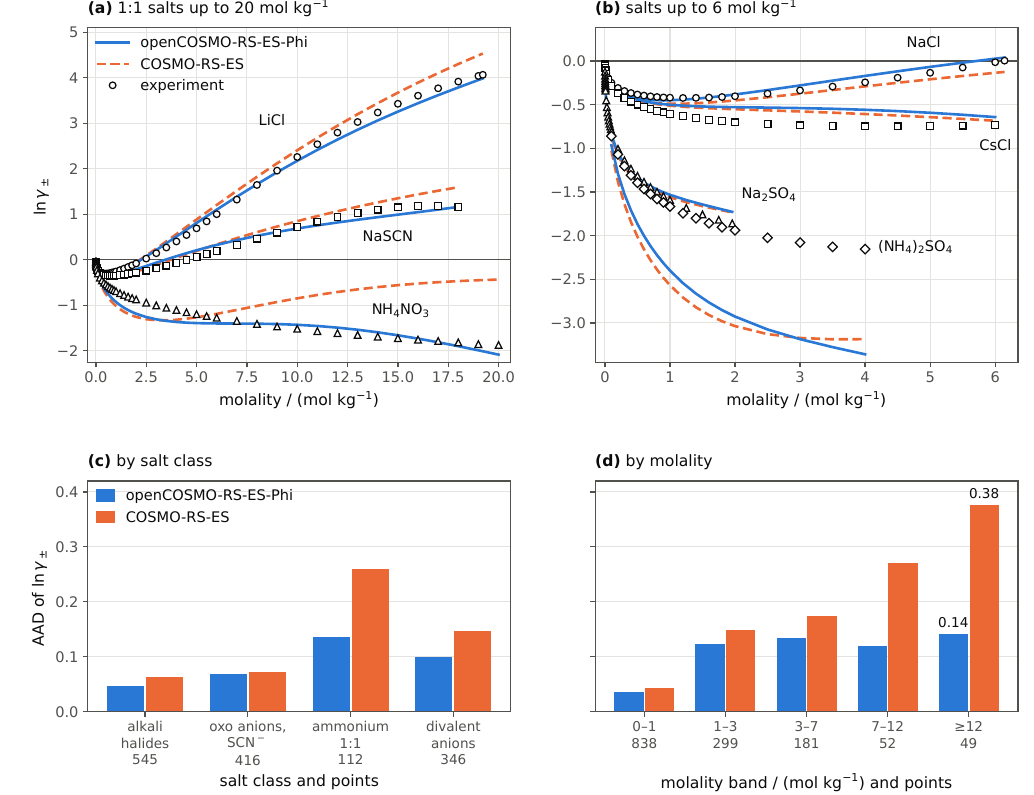}
\caption{MIAC of aqueous salts at 298.15~K, (a) 1:1 salts up to 20~mol~kg$^{-1}$ and (b) salts up to 6~mol~kg$^{-1}$.
Markers: experiment; blue, solid: \ESPhi{}; orange, dashed: COSMO-RS-ES.
AAD of $\ln\gamma_\pm$ over all 51 training salts (c) by salt class and (d) by molality band, with the number of points below each bar.
Ammonium sulfate and ammonium hydrogen phosphate count as salts of divalent anions.}
\label{fig:aqueous-miac}
\end{figure}

Both models are fitted to the MIAC of the 51 aqueous training salts up to 20~mol~kg$^{-1}$ (Figure~\ref{fig:aqueous-miac}, and Figure~\ref{SI-fig:aqueous-parity} of the Supporting Information for all points).
\ESPhi{} reaches an AAD of 0.073 against 0.101 for COSMO-RS-ES and is better for 29 of the 51 salts.

At low concentrations the two models nearly coincide (Figure~\ref{fig:aqueous-miac}d), since both reduce to the Debye--H\"uckel limiting law of the salt-free solvent \citep{Hist12Debye1}.

\ESPhi{} is better in all four salt classes, most for the ammonium salts and the salts of divalent anions (Figure~\ref{fig:aqueous-miac}c).
Many errors of COSMO-RS-ES shift a whole curve, as for KF, NaNO$_3$ and Na$_2$S$_2$O$_3$, because its interaction energies are defined per ion class \citep{gerlachDevelopmentCOSMORSBased2018,gonzalezdecastillaRevisionExtensionCOSMORS2025}, so one salt cannot be corrected without moving others.
The ion volumes of \ESPhi{} are fitted per ion and largely remove these offsets.
Since one cation volume must serve halides and oxo anions alike, a few salts such as KNO$_3$ and CsCl become worse (Figure~\ref{fig:aqueous-miac}b).

The gap widens with molality and reaches 0.141 against 0.377 above 12~mol~kg$^{-1}$.
At these molalities the liquid is no longer incompressible, since the electrostatic pressure $-(\partial A^\mathrm{LR}/\partial V)_{T,\mathbf{n}}$ of \ESPhi{} reaches 156~MPa for concentrated LiCl, and only \ESPhi{} lets the volume respond.
The steep rise of $\ln\gamma_\pm$ then results from two competing contributions.
The decrement keeps the long-range term falling, while the ion cavities and the negative hole volumes (Section~\ref{sec:adjusted}) fill the free volume, so that the repulsion rises steeply.
Physically, the rise reflects water that is bound in the hydration shells and lost as solvent \citep{stokesIonicHydrationActivity1948}, which \ESPhi{} captures only effectively.
The remaining error is therefore S-shaped rather than an offset, as for LiCl, whereas COSMO-RS-ES over-predicts throughout (Figure~\ref{fig:aqueous-miac}a).
Because this balance bends only upward, it cannot reproduce the maximum of $\ln\gamma_\pm$ of NaSCN.

Ammonium sulfate is the worst salt of both models (Figure~\ref{fig:aqueous-miac}b).
\ESPhi{} is already far too low at 1~mol~kg$^{-1}$, so the defect lies in the shared interaction energy of NH$_4^+$ and SO$_4^{2-}$ or in the data.

\paragraph{Solution densities}\label{sec:density}

\begin{figure}[htb]
\centering
\includegraphics[width=\textwidth]{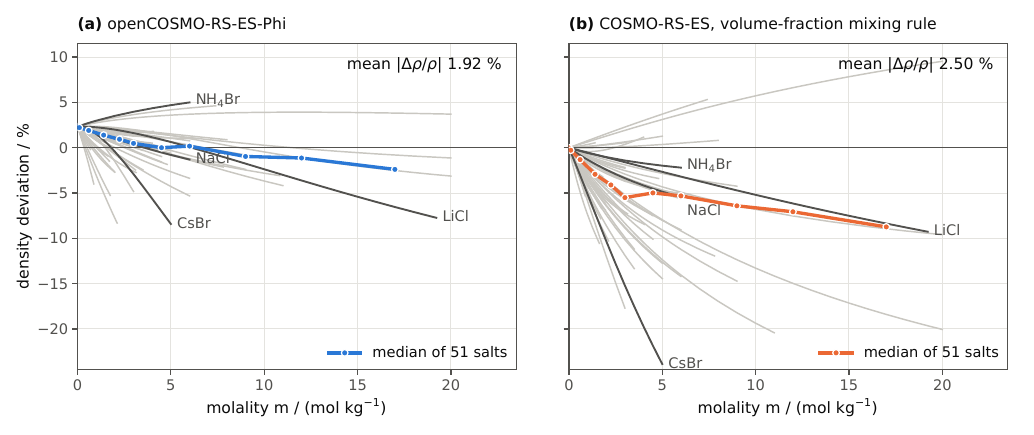}
\caption{Deviation of the density of the 51 aqueous training salts at 298.15~K from the correlations of \citet{apelblatNewTwoparameterEquation2016}, (a) \ESPhi{} and (b) the mixing rule of COSMO-RS-ES.
Gray lines: single salts; colored lines: median.}
\label{fig:density}
\end{figure}

\ESPhi{} yields the density of the solution as part of its state, and its fit includes the densities of the correlations of \citet{apelblatNewTwoparameterEquation2016}.
COSMO-RS-ES has no density of its own.
Its long-range term takes one from a volume-fraction mixture of the pure solvents and a hypothetical molten salt \citep[Eq.~6.56]{gonzalezdecastillaRevisionExtensionCOSMORS2025}, and it is this mixing rule that is compared here.
\ESPhi{} deviates from the correlations by 1.92~\% on average and the mixing rule by 2.50~\% (Figure~\ref{fig:density}).

At low molality all salts start at the $+2.3$~\% offset of the model water, whose parameters are not refitted.
Because the objective contains the absolute density, the fit partly compensates this offset by letting the density rise too weakly with salt, so that the deviation turns negative at high molality.
There it becomes salt-specific, with the ammonium halides too dense and several alkali halides too light.
Hole volumes that do not depend on concentration cannot follow the electrostriction, which fades as the concentration rises and no bulk water is left between the ions \citep{marcusConcentrationDependenceIonic2014,marcusWaterStructureConcentrated2009}.
Universal density correlations are about twice as accurate \citep{nguyenCorrelationsDensitiesAqueous2016a}, and most of the difference is due to the offset of the model water.

The mixing rule of COSMO-RS-ES is exact in the dilute limit but becomes too light with salt, because it ignores the contraction of water around the ions.
Since the melt density depends only weakly on the molar mass, its error also scatters strongly between salts.
Above 1~mol~kg$^{-1}$ \ESPhi{} is therefore the more accurate.

\FloatBarrier									
\paragraph{Temperature dependence}\label{sec:temperature}

\ESPhi{} is extended by two additional parameters (ion--solvent, ion--ion) fitted to the temperature-dependent data from the JESS database, whereas COSMO-RS-ES, without temperature-dependent ion parameters, is applied predictively (Table~\ref{tab:datasets}).
Since the energies enter through Boltzmann factors \citep{klamt_cosmospace_2002}, the fitted ion--solvent weight of $-0.900$, close to $-1$, keeps them nearly constant in units of $RT$.
The factor thus removes the temperature dependence that constant energies would impose rather than describing hydrogen bonding, in line with the ion--water energies fitted in eSAFT-VR Mie \citep{selamThermodynamicModelStrong2018}.

\begin{figure}[!htb]
\centering
\includegraphics[width=\textwidth]{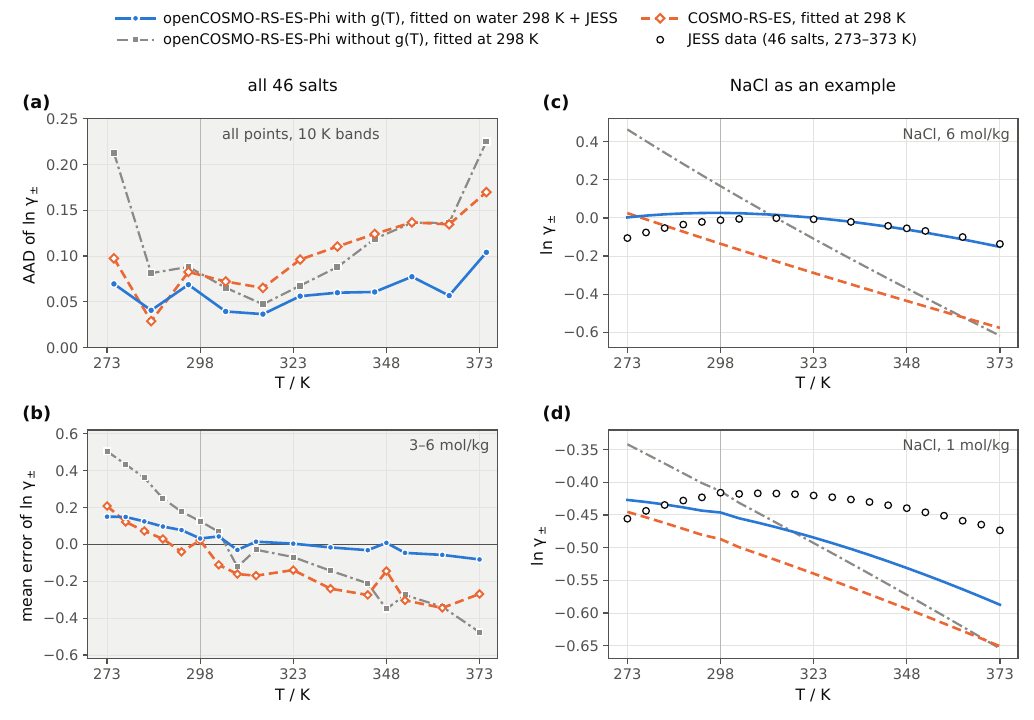}
\caption{MIAC of 46 aqueous salts between 273 and 373~K from JESS \citep{mayJESSJointExpert2010}: (a) AAD in 10~K bands, (b) mean error at 3 to 6~mol~kg$^{-1}$, and NaCl at (c) 6 and (d) 1~mol~kg$^{-1}$.
Blue: \ESPhi{} with $g(T)$; gray, dash-dotted: \ESPhi{} without $g(T)$, fitted at 298.15~K only; orange, dashed: COSMO-RS-ES; circles: data.
The vertical line marks 298.15~K.}
\label{fig:jess}
\end{figure}

Over all JESS points \ESPhi{} reaches an AAD of 0.062 against 0.093 for COSMO-RS-ES, with the largest margin at high temperature and high molality (Figure~\ref{fig:jess}a).
Part of this margin stems from the fit, since \ESPhi{} fitted at 298.15~K alone and without the factor predicts these data less accurately than COSMO-RS-ES.

At high molality, where the data hardly change with temperature, the error of \ESPhi{} still falls with temperature, though less than those of COSMO-RS-ES and of the fit without the factor (Figure~\ref{fig:jess}b).
For NaCl at 6~mol~kg$^{-1}$ the data pass through a flat maximum (Figure~\ref{fig:jess}c), as do those analyzed by \citet{selamThermodynamicModelStrong2018}.
\ESPhi{} reproduces a maximum, although at too low a temperature, whereas COSMO-RS-ES and the fit without the factor fall monotonically, as models without temperature-dependent ion--water interactions do \citep{selamThermodynamicModelStrong2018}.
At 1~mol~kg$^{-1}$ none of the models follows the rise of the data below 298~K (Figure~\ref{fig:jess}d), and neither do other electrolyte equations of state \citep{olsenComparisonsEquationState2023}.

The remaining tilt comes from the long range.
The model water is too dense at low temperature, so its $\varepsilon T$ falls too fast with temperature and the Debye--H\"uckel slope rises too steeply.
Since the ions add no temperature dependence of their own to the permittivity, neither the constant ion volumes nor one factor per group of energies can compensate this.
One factor per group also cannot follow salt-specific trends, such as the opposite tilts of the errors of Li$_2$SO$_4$ and Cs$_2$SO$_4$.

\subsubsection{Mixed solvents}\label{sec:mixed}
\paragraph{Water + cosolvent and organic solvents}\label{sec:mixed-miac}

\begin{figure}[htb]
\centering
\includegraphics[width=\textwidth]{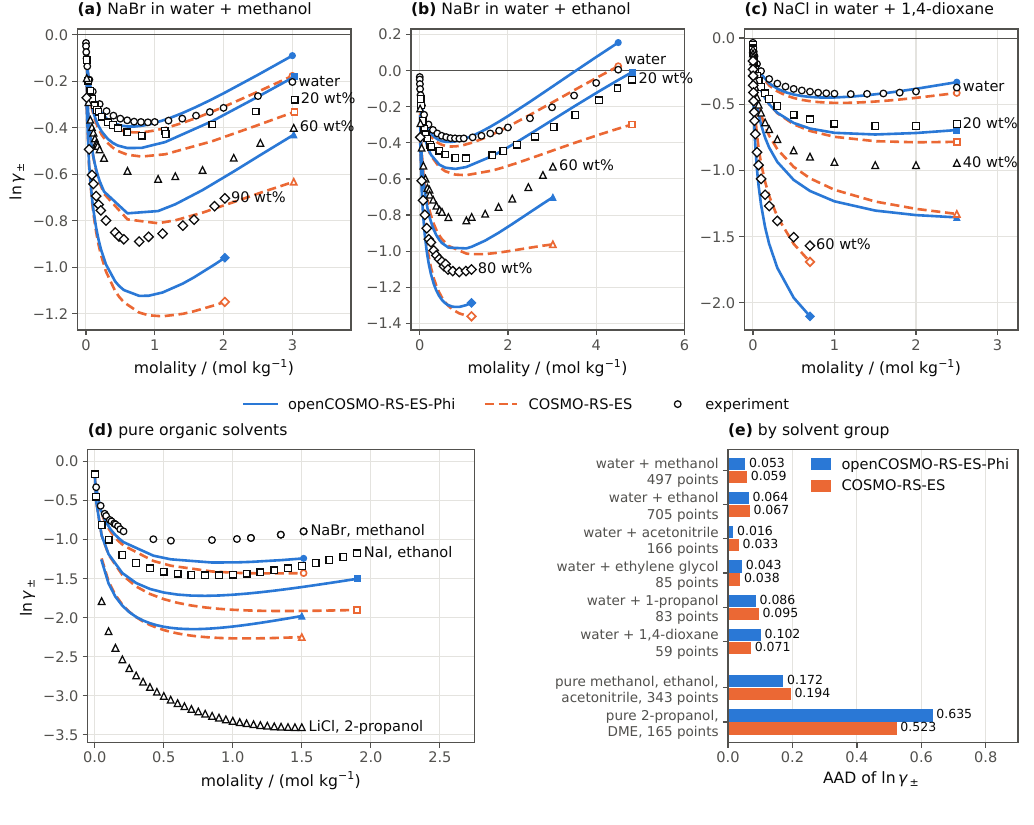}
\caption{Predicted MIAC at 298.15~K of (a) NaBr in water + methanol, (b) NaBr in water + ethanol, (c) NaCl in water + 1,4-dioxane and (d) salts in pure organic solvents.
Markers: experiment; blue, solid: \ESPhi{} (predicted); orange, dashed: COSMO-RS-ES (fitted to these data).
(e) AAD of $\ln\gamma_\pm$ by solvent group.}
\label{fig:mixed-miac}
\end{figure}

In water + cosolvent, \ESPhi{} predicts the MIAC with an AAD of 0.057, as accurately as COSMO-RS-ES reproduces them, with 0.061 (Figure~\ref{fig:mixed-miac}e).
Both models follow the fall of $\ln\gamma_\pm$ as the cosolvent lowers the permittivity, but both place it too low at high cosolvent fractions (Figure~\ref{fig:mixed-miac}a to c).
In pure methanol, ethanol and acetonitrile \ESPhi{} is slightly better than COSMO-RS-ES, whereas in 2-propanol and 1,2-dimethoxyethane both models lie far above the data (Figure~\ref{fig:mixed-miac}d and e).
These are mostly lithium salts, for which ion association, described only implicitly by both models, would lower the activity coefficients.

\paragraph{Water--alcohol mixtures}\label{sec:yang}

The database of \citet{yangBenchmarkDatabaseMixedSolvent2022} holds MIAC and bubble points of alkali halides in water--methanol and water--ethanol over a range of temperatures.
None of these data entered the fit of \ESPhi{}, but its ion--alcohol interaction energies come from COSMO-RS-ES, whose fit contained part of these MIAC.

\begin{figure}[htb]
\centering
\includegraphics[width=\textwidth]{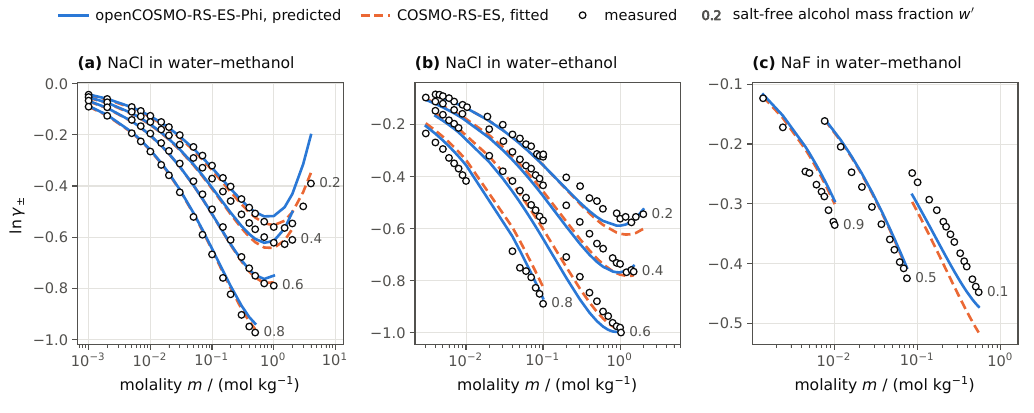}
\caption{Predicted MIAC at 298.15~K of (a) NaCl in water--methanol, (b) NaCl in water--ethanol and (c) NaF in water--methanol, recommended series of \citet{yangBenchmarkDatabaseMixedSolvent2022}, at the alcohol mass fractions given next to each curve.
Blue: \ESPhi{} (predicted); orange, dashed: COSMO-RS-ES (fitted to these data); circles: experiment.}
\label{fig:yang-miac-series}
\end{figure}

The predicted MIAC follow the quality flags of the database, with an AAD of 0.035 on the recommended data and larger errors on the tentative and uncertain data (Figure~\ref{SI-fig:yang-miac-parity} of the Supporting Information).
For the recommended series in Figure~\ref{fig:yang-miac-series}, \ESPhi{} deviates about as little as COSMO-RS-ES, which was fitted to them, and nearly as little as the E-NRTL model that \citet{yangBenchmarkDatabaseMixedSolvent2022} fitted to these data.
The larger errors come from a few sets at high molality and from the cesium salts, for which the model lies above every CsCl and CsBr set, as an ion association not accounted for would cause \citep{mullerCalculationThermodynamicEquilibria2020,bulowEPCSAFTAdvancedPart2021a}.
Since most of these sets could not be verified against other sources, model and data errors cannot be separated.
The largest deviation, for CsI in 75~wt\% methanol \citep{falciolaThermodynamicsAmalgamCells2006}, most likely reflects an experimental error.
There the measured $\gamma_\pm$ rises steeply with salt, whereas that of CsCl and CsBr, measured by the same authors in the same solvent, falls, as expected at low permittivity and as the model predicts.

\begin{figure}[htb]
\centering
\includegraphics[width=\textwidth]{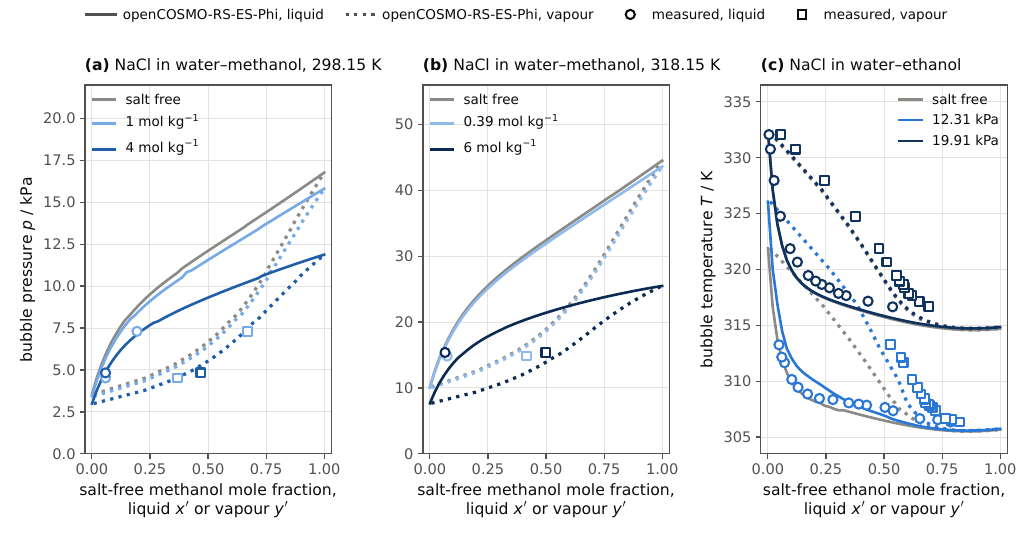}
\caption{Predicted vapor--liquid equilibria of NaCl in water--methanol at (a) 298.15~K \citep{yangVaporLiquidEquilibriaWater1998} and (b) 318.15~K, and (c) in water--ethanol at 12.31 and 19.91~kPa, from the database of \citet{yangBenchmarkDatabaseMixedSolvent2022}.
Lines: \ESPhi{}, solid for the liquid and dotted for the vapor, gray without salt and colored with salt.
Symbols: measured liquid (circles) and vapor (squares), in the color of the line at the same molality.}
\label{fig:yang-vle-diagrams}
\end{figure}

All bubble points converge without any parameter fitted to vapor--liquid data, with mean deviations of 9.8~\% in the bubble pressure and 1.65~K in the bubble temperature (Figure~\ref{SI-fig:yang-vle-parity} of the Supporting Information).
For NaCl in water--methanol and water--ethanol (Figure~\ref{fig:yang-vle-diagrams}), the deviation of the bubble pressures follows that of the vapor pressure of the model water.
At 318.15~K the measured bubble pressure rises slightly with salt, whereas that of the model falls (Figure~\ref{fig:yang-vle-diagrams}b), the wrong sign of the salt effect discussed below.

\begin{figure}[htb]
\centering
\includegraphics[width=\textwidth]{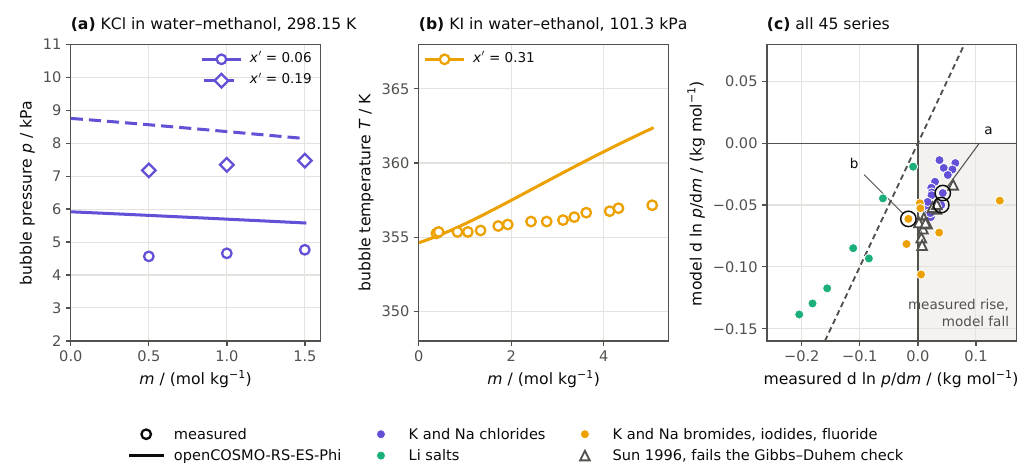}
\caption{Salt effect on the bubble point, predicted by \ESPhi{}.
(a) Bubble pressure of KCl in water--methanol at 298.15~K \citep{yangVaporLiquidEquilibriaWater1998} and (b) bubble temperature of KI in water--ethanol at 101.3~kPa \citep{burnsSaltEffectVaporliquid1979} against molality; symbols: experiment, lines: model.
(c) Change of $\ln p$ with molality, model against experiment, for all 45 series of fixed composition; circled: series of (a) and (b); triangles: data of \citet{sunMolecularThermodynamicsSalt1996}, which fail a Gibbs--Duhem check; dashed line: equal slopes.}
\label{fig:yang-vle}
\end{figure}

The error thus has two parts, the level of the salt-free mixture and the salt effect.
At 298.15~K the salt-free mixture is too volatile, most at the water-rich end (Figure~\ref{fig:yang-vle}a), and this level carries most of the error of the isotherms.
Part of it is pure water, whose vapor pressure the model overestimates by 11~\% at 298.15~K, below the temperature range in which its parameters were fitted \citep{markgrafExtensionOpenCOSMORSFull2026}.
The remaining excess volatility of the alcohol vanishes with temperature and is small at 1~bar \citep{markgrafExtensionOpenCOSMORSFull2026}.

Salt acts on the bubble point in two opposite ways.
In many cases it lowers the activity of water, which lowers the bubble pressure, and it salts out the alcohol, which enriches the vapor in alcohol.
The salting-out is the robust part of the data, since the relative volatility of the alcohol rises with salt in every series of fixed composition, but much less in the model.
Which effect prevails depends on the salt (Figure~\ref{fig:yang-vle}c).
For KCl and NaCl the salting-out prevails, and the measured bubble pressure rises with salt, whereas that of the model falls (Figure~\ref{fig:yang-vle}a).
For KBr and KI in water--ethanol at 1~bar the model overestimates the rise of the bubble temperature \citep{burnsEffectsSaltsHaving1976,burnsSaltEffectVaporliquid1979} (Figure~\ref{fig:yang-vle}b).
Only where the measured bubble pressure falls, for LiCl and concentrated NaBr, does the model have the right sign.
The isobaric data of \citet{sunMolecularThermodynamicsSalt1996} and \citet{booneCorrelationVaporLiquidEquilibrium1976}, whose bubble temperatures fall for all salts, do not contradict this, since the former fail a Gibbs--Duhem check against their own vapor compositions and the latter are flagged uncertain.

The MIAC cannot reveal this error.
Each $\ln\gamma_\pm$ refers to infinite dilution of the salt in the same salt-free mixture (Eq.~\eqref{eq:lngamma}), which removes the standard chemical potential $\mu_\mathrm{s}^\infty(x')$ of the salt at that composition.
The salt effect on the bubble point, however, depends on the change of the alcohol activity relative to that of water, $S=\partial\ln(a_\mathrm{a}/a_\mathrm{w})/\partial m$, which by a Maxwell relation is at low molality
\begin{equation}
S=\frac{\bar M}{RT}\,\frac{\mathrm{d}\mu_{\mathrm{s}}^{\infty}}{\mathrm{d}x'},
\label{eq:yang-maxwell}
\end{equation}
where $\bar M$ is the mean molar mass of the salt-free solvent.
$S$ is thus the composition gradient of the transfer Gibbs energy of the salt, which the MIAC do not constrain.
Measured through the relative volatility, $S$ is clearly positive for the potassium and sodium salts, but close to zero in the model.
Alkali cations, and to a smaller extent halide anions, are preferentially hydrated in water--alcohol mixtures \citep{marcusPreferentialSolvationIons2007,marcusPreferentialSolvationIons2007a} and thereby push the alcohol into the vapor.
In the model this ion-specific solvation is carried only by the interaction energies taken over from COSMO-RS-ES, and the same deficit appears between the pure solvents, where \ESPhi{} transfers NaCl from water to methanol with less than a fifth of the tabulated transfer Gibbs energy \citep{kalidasGibbsEnergiesTransfer2000,marcusGibbsEnergiesTransfer2007}.
Fitted models face the same limit, as the E-NRTL model of \citet{yangBenchmarkDatabaseMixedSolvent2022} captures the rise of the bubble pressure with salt in water--methanol--NaCl only when it is fitted to vapor--liquid data as well.

\paragraph{Salt solubility and salt partitioning}\label{sec:sle-lle}

\begin{figure}[htb]
\centering
\includegraphics[width=\textwidth]{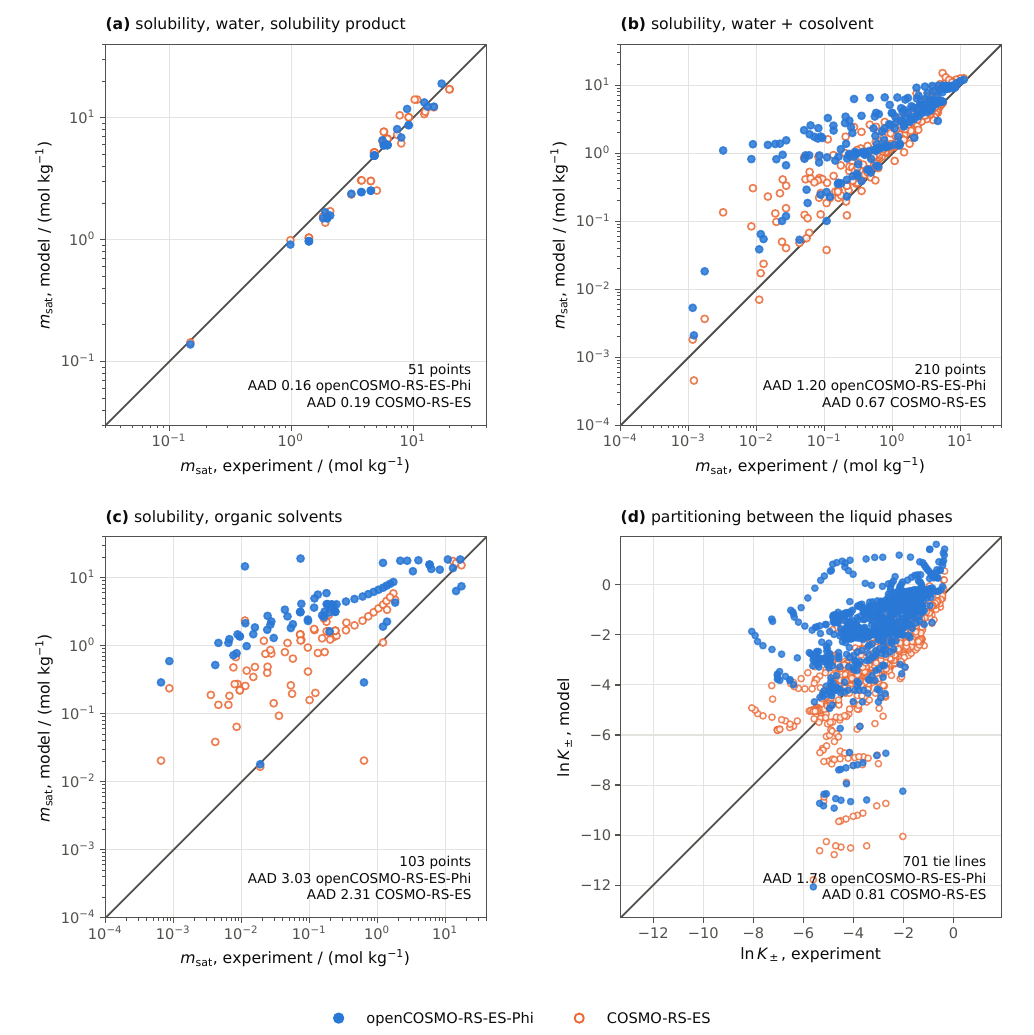}
\caption{Salt solubility (a) in water, (b) in water + cosolvent and (c) in organic solvents, and (d) partition coefficients between an aqueous and an organic-rich phase at 298.15~K; \ESPhi{} (filled) and COSMO-RS-ES (open) against experiment.
In (b) and (c) the measured solubility in water serves as reference; COSMO-RS-ES was fitted to the tie lines in (d).}
\label{fig:sle-lle}
\end{figure}

In water, where the solubility product fixes the solid, both models predict the solubilities at 298.15~K about equally well (Figure~\ref{fig:sle-lle}a and Table~\ref{tab:sle-lle}), apart from some salts that saturate beyond the highest molality of their MIAC data and for which no solubility is found.
With the solubility in water as reference, both models predict the salts too soluble in water + cosolvent and far too soluble in pure organic solvents, \ESPhi{} more so than COSMO-RS-ES (Figure~\ref{fig:sle-lle}b and c).
Both errors reflect the transfer of the ions from water into the organic solvent, which the MIAC cannot constrain (Section~\ref{sec:yang}).

The partitioning of salts between an aqueous and an organic-rich phase tests the same transfer.
\ESPhi{} places too much salt in the organic-rich phase, with an AAD of 1.78 in $\ln K_\pm$, whereas COSMO-RS-ES, fitted to these tie lines, reaches 0.81 (Figure~\ref{fig:sle-lle}d and Table~\ref{tab:sle-lle}).
The error of \ESPhi{} is small in water + methanol, where it is lower than that of COSMO-RS-ES, grows with the chain length of the alcohol and is largest in water + ethyl acetate.
Ammonium sulfate fails in both models, as it does in water (Section~\ref{sec:aqueous-miac}).

\begin{table}[htb]
\centering
\caption{AAD of salt solubility and salt partitioning at 298.15~K, with the mean deviation in parentheses.
The solubility AAD covers only points for which a model finds a solubility below 20~mol~kg$^{-1}$; COSMO-RS-ES was fitted to the tie lines.}
\label{tab:sle-lle}
\small
\begin{tabular}{lrcc}
\hline
 & & \multicolumn{2}{c}{AAD (mean deviation)} \\
 & $n$ & \ESPhi{} & COSMO-RS-ES \\
\hline
\multicolumn{4}{l}{\textbf{Salt solubility, AAD of $\ln m_\mathrm{sat}$}} \\
water, solubility product & 51 & 0.16 ($-$0.12) & 0.19 ($-$0.06) \\
water + cosolvent, water reference & 210 & 1.20 (+1.19) & 0.67 (+0.62) \\
organic solvents, water reference & 103 & 3.03 (+2.96) & 2.31 (+2.21) \\
\addlinespace[1ex]
\multicolumn{4}{l}{\textbf{Salt partitioning, AAD of $\ln K_\pm$}} \\
water + alcohols & 530 & 1.53 (+1.52) & 0.49 (+0.05) \\
water + ketones & 110 & 2.57 (+1.30) & 1.97 ($-$0.85) \\
water + ethyl acetate & 18 & 5.13 (+5.13) & 2.19 (+2.19) \\
water + acetonitrile & 43 & 1.45 ($-$0.10) & 1.14 ($-$0.86) \\
all tie lines & 701 & 1.78 (+1.48) & 0.81 ($-$0.09) \\
\hline
\end{tabular}
\end{table}

Splitting $\ln K_\pm$ into its contributions locates the difference between the models.
The long-range part favors the aqueous phase in both models alike, so the difference lies in the short-range terms that the two models do not share.
The largest systematic one is the dispersion of the cations, fitted to aqueous data, which binds them more strongly to the organic solvents and accounts for most of the overall overestimate.
Without it the systematic overestimate would largely vanish, but a sizeable error would remain, because the rest of the error changes sign between the solvents.
That rest arises in the free-volume terms, where the partitioning is the small difference of a large repulsion and a large attraction, both depending on ion volumes fitted in water.
COSMO-RS-ES absorbs such effects into the parameters it fitted to these tie lines and to transfer Gibbs energies, whereas \ESPhi{} predicts them from aqueous data alone.

\FloatBarrier
\section{Conclusions}\label{sec:conclusions}

This work extended the COSMO-RS-based equation of state \oPhi{} to electrolytes by adding the ion interaction energies and the long-range term of COSMO-RS-ES, which gives \ESPhi{}.
Its ion parameters, the cavity and hole volumes of 20 ions, two dispersion parameters and two temperature weights, were fitted to aqueous data only, while the ion interaction energies are taken over from COSMO-RS-ES, whose fit also contains mixed-solvent data.

Besides the activity coefficients, \ESPhi{} gives access to the density of the solution and its pressure dependence from the same model.

It was found that the dielectric decrement improved the aqueous MIAC, whereas a Born term made every data group worse, presumably because COSMO-RS-ES already describes the solvation of the ions through its short-range term. Therefore the implementation of the dielectric decrement effect is recommended while the Born term is deemed unnecessary.

Compared with COSMO-RS-ES, \ESPhi{}
\begin{itemize}
\item lowers the AAD of the aqueous MIAC, to which both models are fitted, from 0.101 to 0.073, and above 12~mol~kg$^{-1}$, where the volume of the solution responds to the electrostatic pressure, from 0.377 to 0.141,
\item reproduces the solution densities, which enter its fit, to 1.92~\% against 2.50~\% for the mixing rule of COSMO-RS-ES, with ion volumes that contract the solution as electrostriction does,
\item describes the MIAC between 273 and 373~K, to which only \ESPhi{} is fitted, with 0.062 against 0.093, and
\item predicts the MIAC in water + cosolvent, to which only COSMO-RS-ES is fitted, about as accurately, with 0.057 against 0.061.
\end{itemize}
Without vapor--liquid data in any fit, \ESPhi{} predicts the recommended water--alcohol MIAC of \citet{yangBenchmarkDatabaseMixedSolvent2022} with 0.035 and their bubble points with 9.8~\% and 1.65~K.
It predicts, however, the salts too soluble in organic solvents and too strongly partitioned into an organic-rich phase, and it misses the salting-out of the alcohol in the vapor--liquid equilibria.

The remaining deviations, above all in the transfer of ions between solvents, show clearly where the model can be improved, for instance by a transfer contribution \citep{mullerEvaluationRefinementNovel2019a}, a dispersion tied to atomic polarizabilities \citep{grigorashComprehensiveApproachIncorporating2025,grigorashComprehensiveApproachIncorporating2026}, an explicit treatment of ion association \citep{mullerCalculationThermodynamicEquilibria2020} or a permittivity model with dielectric saturation \citep{maribo-mogensenModelingDielectricProperties2013a}.
Since \ESPhi{} is open source and shares its openCOSMO-RS basis with the model for neutral components, each of these extensions can be added and tested directly.
As the first electrolyte equation of state based on openCOSMO-RS, \ESPhi{} thus provides a new basis model on which the predictive description of electrolyte solutions, from aqueous to mixed-solvent systems, can be developed further.

\Contributions{
Leander Raudszus: Methodology, Software, Investigation, Formal analysis, Data curation, Visualization, Writing -- original draft.
Simon M\"uller: Conceptualization, Methodology, Software, Data curation, Supervision, Resources, Project administration, Funding acquisition, Writing -- original draft, Writing -- review \& editing.
}
\Acknowledgments{
Funded by the Deutsche Forschungsgemeinschaft (DFG) -- Project number 503327656.
}
\section*{Data and Code Availability}

The experimental data associated with this study and many more systems are available in the Zenodo repository at \url{https://doi.org/10.5281/zenodo.23668298}.
The source code of \ESPhi{} is available at \url{https://github.com/TUHH-TVT/openCOSMO-RS-ES-Phi_cpp}.

\section*{Declaration of Generative AI Use}

During the preparation of this work the authors used Claude (Anthropic) and ChatGPT (OpenAI) to obtain feedback on the structure of the manuscript, to restructure and copy-edit sections of the text, and to draft the conclusions and the description of the computational workflow.
All scientific content, including the model, the code, the data, the analysis and its interpretation, was developed by the authors.
After using these tools, the authors reviewed and edited the content as needed, verified all numerical results and references, and take full responsibility for the content of the published article.
\EndMatter

@article{apelblatNewTwoparameterEquation2016,
  title = {A New Two-Parameter Equation for Correlation and Prediction of Densities as a Function of Concentration and Temperature in Binary Aqueous Solutions},
  author = {Apelblat, Alexander},
  year = 2016,
  month = jul,
  journal = {Journal of Molecular Liquids},
  volume = {219},
  pages = {313--331},
  issn = {0167-7322},
  doi = {10/f8s4b7},
  urldate = {2018-04-25}
}

@incollection{booneCorrelationVaporLiquidEquilibrium1976,
  title = {The {{Correlation}} of {{Vapor-Liquid Equilibrium Data}} for {{Salt-Containing Systems}}},
  booktitle = {Thermodynamic {{Behavior}} of {{Electrolytes}} in {{Mixed Solvents}}},
  author = {Boone, J. E. and Rousseau, R. W. and Schoenborn, E. M.},
  editor = {Furter, William F.},
  year = 1976,
  month = jun,
  series = {Advances in {{Chemistry}}},
  volume = {155},
  pages = {36--52},
  publisher = {American Chemical Society},
  address = {Washington, DC},
  doi = {10.1021/ba-1976-0155.ch004},
  urldate = {2026-10-25},
  isbn = {978-0-8412-0302-0},
  langid = {english}
}

@article{bornVolumenUndHydratationswaerme1920,
  title = {{Volumen und Hydratationsw\"arme der Ionen}},
  author = {Born, M.},
  year = 1920,
  month = feb,
  journal = {Zeitschrift f\"ur Physik},
  volume = {1},
  number = {1},
  pages = {45--48},
  issn = {1434-6001, 1434-601X},
  doi = {10.1007/BF01881023},
  urldate = {2026-10-18},
  copyright = {http://www.springer.com/tdm},
  langid = {ngerman}
}

@article{buchnerDielectricRelaxationAqueous1999,
  title = {Dielectric {{Relaxation}} of {{Aqueous NaCl Solutions}}},
  author = {Buchner, Richard and Hefter, Glenn T. and May, Peter M.},
  year = 1999,
  month = jan,
  journal = {The Journal of Physical Chemistry A},
  volume = {103},
  number = {1},
  pages = {1--9},
  issn = {1089-5639, 1520-5215},
  doi = {10.1021/jp982977k},
  urldate = {2026-10-18},
  langid = {english}
}

@article{bulowEPCSAFTAdvancedPart2021a,
  title = {{{ePC-SAFT}} Advanced -- {{Part II}}: {{Application}} to {{Salt Solubility}} in {{Ionic}} and {{Organic Solvents}} and the {{Impact}} of {{Ion Pairing}}},
  shorttitle = {{{ePC-SAFT}} Advanced -- {{Part II}}},
  author = {B{\"u}low, Mark and Ascani, Moreno and Held, Christoph},
  year = 2021,
  month = jun,
  journal = {Fluid Phase Equilibria},
  volume = {537},
  pages = {112989},
  issn = {0378-3812},
  doi = {10.1016/j.fluid.2021.112989},
  urldate = {2021-03-12},
  langid = {english}
}

@article{bulowEPCSAFTAdvancedPart2021b,
  title = {{{ePC-SAFT}} Advanced - {{Part I}}: {{Physical}} Meaning of Including a Concentration-Dependent Dielectric Constant in the Born Term and in the {{Debye-H\"uckel}} Theory},
  shorttitle = {{{ePC-SAFT}} Advanced - {{Part I}}},
  author = {B{\"u}low, Mark and Ascani, Moreno and Held, Christoph},
  year = 2021,
  month = may,
  journal = {Fluid Phase Equilibria},
  volume = {535},
  pages = {112967},
  issn = {03783812},
  doi = {10.1016/j.fluid.2021.112967},
  urldate = {2026-08-16},
  langid = {english}
}

@incollection{burnsEffectsSaltsHaving1976,
  title = {Effects of {{Salts Having Large Organic Ions}} on {{Vapor-Liquid Equilibrium}}},
  booktitle = {Thermodynamic {{Behavior}} of {{Electrolytes}} in {{Mixed Solvents}}},
  author = {Burns, John A. and Furter, William F.},
  editor = {Furter, William F.},
  year = 1976,
  month = jun,
  series = {Advances in {{Chemistry}}},
  volume = {155},
  pages = {99--127},
  publisher = {American Chemical Society},
  address = {Washington, DC},
  doi = {10.1021/ba-1976-0155.ch008},
  urldate = {2026-10-25},
  isbn = {978-0-8412-0302-0},
  langid = {english}
}

@article{camerettiModelingAqueousElectrolyte2005,
  title = {Modeling of {{Aqueous Electrolyte Solutions}} with {{Perturbed-Chain Statistical Associated Fluid Theory}}},
  author = {Cameretti, Luca F. and Sadowski, Gabriele and Mollerup, J{\o}rgen M.},
  year = 2005,
  month = apr,
  journal = {Industrial \& Engineering Chemistry Research},
  volume = {44},
  number = {9},
  pages = {3355--3362},
  issn = {0888-5885, 1520-5045},
  doi = {10.1021/ie0488142},
  urldate = {2014-05-14},
  langid = {english}
}

@article{chenGeneralizedElectrolyteNRTLModel2004,
  title = {Generalized Electrolyte-{{NRTL}} Model for Mixed-Solvent Electrolyte Systems},
  author = {Chen, Chau-Chyun and Song, Yuhua},
  year = 2004,
  journal = {AIChE Journal},
  volume = {50},
  number = {8},
  pages = {1928--1941},
  issn = {1547-5905},
  doi = {10.1002/aic.10151},
  urldate = {2026-07-26},
  copyright = {Copyright \copyright{} 2004 American Institute of Chemical Engineers (AIChE)},
  langid = {english}
}

@article{chenLocalCompositionModel1982,
  title = {Local Composition Model for Excess {{Gibbs}} Energy of Electrolyte Systems. {{Part I}}: {{Single}} Solvent, Single Completely Dissociated Electrolyte Systems},
  shorttitle = {Local Composition Model for Excess {{Gibbs}} Energy of Electrolyte Systems. {{Part I}}},
  author = {Chen, Chau-Chyun and Britt, H. I. and Boston, J. F. and Evans, L. B.},
  year = 1982,
  journal = {AIChE Journal},
  volume = {28},
  number = {4},
  pages = {588--596},
  issn = {1547-5905},
  doi = {10.1002/aic.690280410},
  urldate = {2026-07-26},
  langid = {english}
}

@article{chenLocalCompositionModel1986,
  title = {A Local Composition Model for the Excess {{Gibbs}} Energy of Aqueous Electrolyte Systems},
  author = {Chen, Chau-Chyun and Evans, L. B.},
  year = 1986,
  month = mar,
  journal = {AIChE Journal},
  volume = {32},
  number = {3},
  pages = {444--454},
  issn = {0001-1541, 1547-5905},
  doi = {10.1002/aic.690320311},
  urldate = {2026-07-26},
  copyright = {http://onlinelibrary.wiley.com/termsAndConditions\#vor},
  langid = {english}
}

@article{dep.soaresPairwiseSurfaceContact2019,
  title = {A Pairwise Surface Contact Equation of State: {{COSMO-SAC-Phi}}},
  shorttitle = {A Pairwise Surface Contact Equation of State},
  author = {De P. Soares, Rafael and Balad{\~a}o, Luis F. and Staudt, Paula B.},
  year = 2019,
  month = jun,
  journal = {Fluid Phase Equilibria},
  volume = {488},
  pages = {13--26},
  issn = {03783812},
  doi = {10.1016/j.fluid.2019.01.015},
  urldate = {2026-02-12},
  langid = {english}
}

@article{eriksenDevelopmentIntermolecularPotential2016,
  title = {Development of Intermolecular Potential Models for Electrolyte Solutions Using an Electrolyte {{SAFT-VR Mie}} Equation of State},
  author = {Eriksen, Daniel K. and Lazarou, Georgia and Galindo, Amparo and Jackson, George and Adjiman, Claire S. and Haslam, Andrew J.},
  year = 2016,
  month = sep,
  journal = {Molecular Physics},
  volume = {114},
  number = {18},
  pages = {2724--2749},
  issn = {0026-8976, 1362-3028},
  doi = {10.1080/00268976.2016.1236221},
  urldate = {2017-03-12},
  langid = {english}
}

@article{falciolaThermodynamicsAmalgamCells2006,
  title = {Thermodynamics of the Amalgam Cells \textbraceleft{{Cs}}-amalgam\textbar{{CsX}} (m)\textbar{{AgX}}\textbar{{Ag}}\textbraceright{} ({{X}} = {{Cl}}, {{Br}}, {{I}}) and Primary Medium Effects in (Methanol + Water), (Acetonitrile + Water), and (1,4-Dioxane + Water) Solvent Mixtures},
  author = {Falciola, Luigi and Longoni, Giorgio and Mussini, Patrizia R. and Mussini, Torquato},
  year = 2006,
  month = jun,
  journal = {The Journal of Chemical Thermodynamics},
  volume = {38},
  number = {6},
  pages = {788--798},
  issn = {00219614},
  doi = {10.1016/j.jct.2005.08.014},
  urldate = {2026-10-25},
  langid = {english}
}

@article{furstRepresentationExcessProperties1993,
  title = {{Representation of excess properties of electrolyte solutions using a new equation of state}},
  author = {F{\"u}rst, W. and Renon, H.},
  year = 1993,
  journal = {AIChE Journal},
  volume = {39},
  number = {2},
  pages = {335--343},
  issn = {1547-5905},
  doi = {10.1002/aic.690390213},
  urldate = {2019-12-06},
  copyright = {Copyright \copyright{} 1993 American Institute of Chemical Engineers},
  langid = {english}
}

@article{galindoSAFTVREPhaseBehavior1999b,
  title = {{{SAFT-VRE}}: {{Phase Behavior}} of {{Electrolyte Solutions}} with the {{Statistical Associating Fluid Theory}} for {{Potentials}} of {{Variable Range}}},
  shorttitle = {{{SAFT-VRE}}},
  author = {Galindo, Amparo and {Gil-Villegas}, Alejandro and Jackson, George and Burgess, Andrew N.},
  year = 1999,
  month = nov,
  journal = {The Journal of Physical Chemistry B},
  volume = {103},
  number = {46},
  pages = {10272--10281},
  issn = {1520-6106, 1520-5207},
  doi = {10.1021/jp991959f},
  urldate = {2026-07-26},
  langid = {english}
}

@article{gerlachDevelopmentCOSMORSBased2018,
  title = {Development of a {{COSMO}}-{{RS}} Based Model for the Calculation of Phase Equilibria in Electrolyte Systems},
  author = {Gerlach, Thomas and M{\"u}ller, Simon and Smirnova, Irina},
  year = 2018,
  month = jan,
  journal = {AIChE Journal},
  volume = {64},
  number = {1},
  pages = {272--285},
  issn = {0001-1541, 1547-5905},
  doi = {10.1002/aic.15875},
  urldate = {2026-07-05},
  copyright = {http://onlinelibrary.wiley.com/termsAndConditions\#vor},
  langid = {english}
}

@article{gerlachOpenSourceCOSMORS2022,
  title = {An Open Source {{COSMO-RS}} Implementation and Parameterization Supporting the Efficient Implementation of Multiple Segment Descriptors},
  author = {Gerlach, Thomas and M{\"u}ller, Simon and De Castilla, Andr{\'e}s Gonz{\'a}lez and Smirnova, Irina},
  year = 2022,
  month = sep,
  journal = {Fluid Phase Equilibria},
  volume = {560},
  pages = {113472},
  issn = {03783812},
  doi = {10.1016/j.fluid.2022.113472},
  urldate = {2026-02-12},
  langid = {english}
}

@article{gonzalezdecastillaAnalogyRestrictedPrimitive2021,
  title = {On the Analogy between the Restricted Primitive Model and Capacitor Circuits: {{Semi}}-Empirical Alternatives for over- and Underscreening in the Calculation of Mean Ionic Activity Coefficients},
  shorttitle = {On the Analogy between the Restricted Primitive Model and Capacitor Circuits},
  author = {Gonz{\'a}lez De Castilla, Andr{\'e}s and M{\"u}ller, Simon and Smirnova, Irina},
  year = 2021,
  month = mar,
  journal = {Journal of Molecular Liquids},
  volume = {326},
  pages = {115204},
  issn = {01677322},
  doi = {10.1016/j.molliq.2020.115204},
  urldate = {2026-03-09},
  langid = {english}
}

@article{gonzalezdecastillaAnalogyRestrictedPrimitive2022,
  title = {On the Analogy between the Restricted Primitive Model and Capacitor Circuits. {{Part II}}: {{A}} Generalized {{Gibbs-Duhem}} Consistent Extension of the {{Pitzer-Debye-H\"uckel}} Term with Corrections for Low and Variable Relative Permittivity},
  shorttitle = {On the Analogy between the Restricted Primitive Model and Capacitor Circuits. {{Part II}}},
  author = {Gonz{\'a}lez De Castilla, Andr{\'e}s and M{\"u}ller, Simon and Smirnova, Irina},
  year = 2022,
  month = aug,
  journal = {Journal of Molecular Liquids},
  volume = {360},
  pages = {119398},
  issn = {01677322},
  doi = {10.1016/j.molliq.2022.119398},
  urldate = {2026-03-09},
  langid = {english}
}

@phdthesis{gonzalezdecastillaRevisionExtensionCOSMORS2025,
  title = {Revision and Extension of a {{COSMO-RS}} Based Model for Electrolyte Systems: From Infinite Dilution to the Fused Salt State},
  shorttitle = {Revision and Extension of a {{COSMO-RS}} Based Model for Electrolyte Systems},
  author = {{Gonz{\'a}lez de Castilla}, Andr{\'e}s},
  year = 2025,
  school = {Hamburg University of Technology},
  address = {Hamburg},
  doi = {10.15480/882.16280},
  urldate = {2026-03-09},
  archiveprefix = {TUHH Universit\"atsbibliothek},
  collaborator = {{TUHH Universit\"atsbibliothek} and Klamt, Andreas},
  copyright = {Creative Commons Attribution 4.0 International},
  langid = {english}
}

@article{haggisDielectricPropertiesWater1952,
  title = {The {{Dielectric Properties}} of {{Water}} in {{Solutions}}},
  author = {Haggis, G. H. and Hasted, J. B. and Buchanan, T. J.},
  year = 1952,
  month = sep,
  journal = {The Journal of Chemical Physics},
  volume = {20},
  number = {9},
  pages = {1452--1465},
  issn = {0021-9606, 1089-7690},
  doi = {10.1063/1.1700780},
  urldate = {2026-10-25},
  langid = {english}
}

@article{harrisDielectricPropertiesAqueous1957,
  title = {Dielectric {{Properties}} of {{Aqueous Ionic Solutions}} at {{Microwave Frequencies}}},
  author = {Harris, Frank E. and O'Konski, Chester T.},
  year = 1957,
  month = mar,
  journal = {The Journal of Physical Chemistry},
  volume = {61},
  number = {3},
  pages = {310--319},
  issn = {0022-3654, 1541-5740},
  doi = {10.1021/j150549a009},
  urldate = {2026-10-25},
  langid = {english}
}

@article{hastedDielectricPropertiesAqueous1948,
  title = {Dielectric {{Properties}} of {{Aqueous Ionic Solutions}}. {{Parts I}} and {{II}}},
  author = {Hasted, J. B. and Ritson, D. M. and Collie, C. H.},
  year = 1948,
  month = jan,
  journal = {The Journal of Chemical Physics},
  volume = {16},
  number = {1},
  pages = {1--21},
  issn = {0021-9606, 1089-7690},
  doi = {10.1063/1.1746645},
  urldate = {2026-10-18},
  langid = {english}
}

@article{hastedDielectricPropertiesAqueous1958,
  title = {Dielectric {{Properties}} of {{Aqueous}} and {{Alcoholic Electrolytic Solutions}}},
  author = {Hasted, J. B. and Roderick, G. W.},
  year = 1958,
  month = jul,
  journal = {The Journal of Chemical Physics},
  volume = {29},
  number = {1},
  pages = {17--26},
  issn = {0021-9606, 1089-7690},
  doi = {10.1063/1.1744418},
  urldate = {2026-10-25},
  langid = {english}
}

@article{heldEPCSAFTRevised2014,
  title = {{{ePC-SAFT}} Revised},
  author = {Held, Christoph and Reschke, Thomas and Mohammad, Sultan and Luza, Armando and Sadowski, Gabriele},
  year = 2014,
  month = dec,
  journal = {Chemical Engineering Research and Design},
  volume = {92},
  number = {12},
  pages = {2884--2897},
  issn = {02638762},
  doi = {10.1016/j.cherd.2014.05.017},
  urldate = {2015-04-19},
  langid = {english}
}

@article{heldModelingAqueousElectrolyte2008,
  title = {Modeling Aqueous Electrolyte Solutions},
  author = {Held, Christoph and Cameretti, Luca F. and Sadowski, Gabriele},
  year = 2008,
  month = aug,
  journal = {Fluid Phase Equilibria},
  volume = {270},
  number = {1-2},
  pages = {87--96},
  issn = {03783812},
  doi = {10.1016/j.fluid.2008.06.010},
  urldate = {2013-12-07}
}

@article{heldThermodynamicGEModels2020,
  title = {Thermodynamic {\emph{g}}{{{\textsuperscript{E}}}} {{Models}} and {{Equations}} of {{State}} for {{Electrolytes}} in a {{Water-Poor Medium}}: {{A Review}}},
  shorttitle = {Thermodynamic {\emph{g}}{{{\textsuperscript{E}}}} {{Models}} and {{Equations}} of {{State}} for {{Electrolytes}} in a {{Water-Poor Medium}}},
  author = {Held, Christoph},
  year = 2020,
  month = nov,
  journal = {Journal of Chemical \& Engineering Data},
  volume = {65},
  number = {11},
  pages = {5073--5082},
  issn = {0021-9568, 1520-5134},
  doi = {10.1021/acs.jced.0c00812},
  urldate = {2026-04-09},
  copyright = {https://doi.org/10.15223/policy-029},
  langid = {english}
}

@article{Hist12Debye1,
  title = {Zur {{Theorie}} Der {{Elektrolyte}}. {{I}}. {{Gefrierpunktserniedrigung}} Und Verwandte {{Erscheinungen}}},
  author = {Debye, P. and H{\"u}ckel, E.},
  year = 1923,
  journal = {Physikalische Zeitschrift},
  volume = {24},
  number = {9},
  pages = {185--206},
  langid = {ngerman}
}

@article{hsiehPredictiveModelExcess2011,
  title = {A Predictive Model for the Excess Gibbs Free Energy of Fully Dissociated Electrolyte Solutions},
  author = {Hsieh, Meng-Ting and Lin, Shiang-Tai},
  year = 2011,
  month = apr,
  journal = {AIChE Journal},
  volume = {57},
  number = {4},
  pages = {1061--1074},
  issn = {0001-1541, 1547-5905},
  doi = {10.1002/aic.12325},
  urldate = {2026-07-26},
  copyright = {http://onlinelibrary.wiley.com/termsAndConditions\#vor},
  langid = {english}
}

@article{hubbardDielectricDispersionDielectric1977,
  title = {Dielectric Dispersion and Dielectric Friction in Electrolyte Solutions. {{I}}.},
  author = {Hubbard, J. and Onsager, L.},
  year = 1977,
  month = dec,
  journal = {The Journal of Chemical Physics},
  volume = {67},
  number = {11},
  pages = {4850--4857},
  issn = {0021-9606, 1089-7690},
  doi = {10.1063/1.434664},
  urldate = {2026-10-25},
  langid = {english}
}

@article{ingramExtensionCOSMORSMonoatomic2012,
  title = {Extension of {{COSMO-RS}} for Monoatomic Electrolytes: {{Modeling}} of Liquid--Liquid Equilibria in Presence of Salts},
  shorttitle = {Extension of {{COSMO-RS}} for Monoatomic Electrolytes},
  author = {Ingram, Thomas and Gerlach, Thomas and Mehling, Tanja and Smirnova, Irina},
  year = 2012,
  month = jan,
  journal = {Fluid Phase Equilibria},
  volume = {314},
  pages = {29--37},
  issn = {03783812},
  doi = {10.1016/j.fluid.2011.09.021},
  urldate = {2026-07-26},
  copyright = {https://www.elsevier.com/tdm/userlicense/1.0/},
  langid = {english}
}

@article{kalidasGibbsEnergiesTransfer2000,
  title = {Gibbs {{Energies}} of {{Transfer}} of {{Cations}} from {{Water}} to {{Mixed Aqueous Organic Solvents}}},
  author = {Kalidas, C. and Hefter, Glenn and Marcus, Yizhak},
  year = 2000,
  month = mar,
  journal = {Chemical Reviews},
  volume = {100},
  number = {3},
  pages = {819--852},
  issn = {0009-2665, 1520-6890},
  doi = {10.1021/cr980144k},
  urldate = {2026-10-18},
  langid = {english}
}

@article{klamtConductorlikeScreeningModel1995,
  title = {Conductor-like {{Screening Model}} for {{Real Solvents}}: {{A New Approach}} to the {{Quantitative Calculation}} of {{Solvation Phenomena}}},
  shorttitle = {Conductor-like {{Screening Model}} for {{Real Solvents}}},
  author = {Klamt, Andreas},
  year = 1995,
  month = feb,
  journal = {The Journal of Physical Chemistry},
  volume = {99},
  number = {7},
  pages = {2224--2235},
  issn = {0022-3654, 1541-5740},
  doi = {10.1021/j100007a062},
  urldate = {2026-02-13},
  langid = {english}
}

@article{klamtCOSMOCOSMORSSolvation2018,
  title = {The {{{\textsc{COSMO}}}} and {{{\textsc{COSMO}}}}{\textsc{-}}{{{\textsc{RS}}}} Solvation Models},
  shorttitle = {The},
  author = {Klamt, Andreas},
  year = 2018,
  month = jan,
  journal = {WIREs Computational Molecular Science},
  volume = {8},
  number = {1},
  pages = {e1338},
  issn = {1759-0876, 1759-0884},
  doi = {10.1002/wcms.1338},
  urldate = {2026-10-18},
  copyright = {http://onlinelibrary.wiley.com/termsAndConditions\#vor},
  langid = {english}
}

@article{klamtCOSMONewApproach1993,
  title = {{{COSMO}}: A New Approach to Dielectric Screening in Solvents with Explicit Expressions for the Screening Energy and Its Gradient},
  shorttitle = {{{COSMO}}},
  author = {Klamt, A. and Sch{\"u}{\"u}rmann, G.},
  year = 1993,
  journal = {J. Chem. Soc., Perkin Trans. 2},
  number = {5},
  pages = {799--805},
  issn = {0300-9580, 1364-5471},
  doi = {10.1039/P29930000799},
  urldate = {2026-10-18},
  langid = {english}
}

@article{klamtRefinementParametrizationCOSMORS1998,
  title = {Refinement and {{Parametrization}} of {{COSMO-RS}}},
  author = {Klamt, Andreas and Jonas, Volker and B{\"u}rger, Thorsten and Lohrenz, John C. W.},
  year = 1998,
  month = jun,
  journal = {The Journal of Physical Chemistry A},
  volume = {102},
  number = {26},
  pages = {5074--5085},
  issn = {1089-5639, 1520-5215},
  doi = {10.1021/jp980017s},
  urldate = {2026-02-12},
  langid = {english}
}

@article{kontogeorgisDebyeHuckelTheoryIts2018,
  title = {The {{Debye-H\"uckel}} Theory and Its Importance in Modeling Electrolyte Solutions},
  author = {Kontogeorgis, Georgios M. and {Maribo-Mogensen}, Bj{\o}rn and Thomsen, Kaj},
  year = 2018,
  month = apr,
  journal = {Fluid Phase Equilibria},
  volume = {462},
  pages = {130--152},
  issn = {0378-3812},
  doi = {10.1016/j.fluid.2018.01.004},
  urldate = {2021-10-28},
  langid = {english}
}

@article{kontogeorgisReviewElectrolyteEquations2022,
  title = {A {{Review}} of {{Electrolyte Equations}} of {{State}} with {{Emphasis}} on {{Those Based}} on {{Cubic}} and {{Cubic-Plus-Association}} ({{CPA}}) {{Models}}},
  author = {Kontogeorgis, Georgios M. and Schlaikjer, Anders and Olsen, Martin Due and {Maribo-Mogensen}, Bj{\o}rn and Thomsen, Kaj and {von Solms}, Nicolas and Liang, Xiaodong},
  year = 2022,
  month = feb,
  journal = {International Journal of Thermophysics},
  volume = {43},
  number = {4},
  pages = {54},
  issn = {1572-9567},
  doi = {10.1007/s10765-022-02976-4},
  urldate = {2022-02-12},
  langid = {english}
}

@article{latimerFreeEnergyHydration1939,
  title = {The {{Free Energy}} of {{Hydration}} of {{Gaseous Ions}}, and the {{Absolute Potential}} of the {{Normal Calomel Electrode}}},
  author = {Latimer, Wendell M. and Pitzer, Kenneth S. and Slansky, Cyril M.},
  year = 1939,
  month = feb,
  journal = {The Journal of Chemical Physics},
  volume = {7},
  number = {2},
  pages = {108--111},
  issn = {0021-9606, 1089-7690},
  doi = {10.1063/1.1750387},
  urldate = {2026-10-18},
  langid = {english}
}

@article{liGEModelSingle1994,
  title = {A {\emph{g}}{{E}} Model for Single and Mixed Solvent Electrolyte Systems},
  author = {Li, Jiding and Polka, Hans-Martin and Gmehling, J{\"u}rgen},
  year = 1994,
  month = mar,
  journal = {Fluid Phase Equilibria},
  volume = {94},
  pages = {89--114},
  issn = {0378-3812},
  doi = {10.1016/0378-3812(94)87052-7},
  urldate = {2026-07-26}
}

@article{linPrioriPhaseEquilibrium2002,
  title = {A {{Priori Phase Equilibrium Prediction}} from a {{Segment Contribution Solvation Model}}},
  author = {Lin, Shiang-Tai and Sandler, Stanley I.},
  year = 2002,
  month = mar,
  journal = {Industrial \& Engineering Chemistry Research},
  volume = {41},
  number = {5},
  pages = {899--913},
  issn = {0888-5885, 1520-5045},
  doi = {10.1021/ie001047w},
  urldate = {2026-07-26},
  langid = {english}
}

@article{marcusConcentrationDependenceIonic2014,
  title = {Concentration {{Dependence}} of {{Ionic Hydration Numbers}}},
  author = {Marcus, Yizhak},
  year = 2014,
  month = sep,
  journal = {The Journal of Physical Chemistry B},
  volume = {118},
  number = {35},
  pages = {10471--10476},
  issn = {1520-6106, 1520-5207},
  doi = {10.1021/jp5039255},
  urldate = {2026-10-19},
  langid = {english}
}

@article{marcusElectrostrictionElectrolyteSolutions2011a,
  title = {Electrostriction in {{Electrolyte Solutions}}},
  author = {Marcus, Yizhak},
  year = 2011,
  month = apr,
  journal = {Chemical Reviews},
  volume = {111},
  number = {4},
  pages = {2761--2783},
  issn = {0009-2665, 1520-6890},
  doi = {10.1021/cr100130d},
  urldate = {2026-10-18},
  langid = {english}
}

@article{marcusGibbsEnergiesTransfer2007,
  title = {Gibbs {{Energies}} of {{Transfer}} of {{Anions}} from {{Water}} to {{Mixed Aqueous Organic Solvents}}},
  author = {Marcus, Yizhak},
  year = 2007,
  month = sep,
  journal = {Chemical Reviews},
  volume = {107},
  number = {9},
  pages = {3880--3897},
  issn = {0009-2665, 1520-6890},
  doi = {10.1021/cr068045r},
  urldate = {2026-10-18},
  langid = {english}
}

@article{marcusPreferentialSolvationIons2007,
  title = {Preferential {{Solvation}} of {{Ions}} in {{Mixed Solvents}}. 5. {{The Alkali Metal}}, {{Silver}}, and {{Thallium}}({{I}}) {{Cations}} in {{Aqueous Organic Solvents According}} to the {{Inverse Kirkwood-Buff Integral}} ({{IKBI}}) {{Approach}}},
  author = {Marcus, Yizhak},
  year = 2007,
  month = nov,
  journal = {Journal of Solution Chemistry},
  volume = {36},
  number = {11-12},
  pages = {1385--1399},
  issn = {0095-9782, 1572-8927},
  doi = {10.1007/s10953-007-9196-4},
  urldate = {2026-10-18},
  copyright = {http://www.springer.com/tdm},
  langid = {english}
}

@article{marcusPreferentialSolvationIons2007a,
  title = {Preferential Solvation of Ions in Mixed Solvents. 6: {{Univalent}} Anions in Aqueous Organic Solvents According to the Inverse {{Kirkwood}}--{{Buff}} Integral ({{IKBI}}) Approach},
  shorttitle = {Preferential Solvation of Ions in Mixed Solvents. 6},
  author = {Marcus, Yizhak},
  year = 2007,
  month = oct,
  journal = {The Journal of Chemical Thermodynamics},
  volume = {39},
  number = {10},
  pages = {1338--1345},
  issn = {00219614},
  doi = {10.1016/j.jct.2007.04.003},
  urldate = {2026-10-18},
  langid = {english}
}

@article{marcusStandardPartialMolar2004,
  title = {Standard {{Partial Molar Volumes}} of {{Electrolytes}} and {{Ions}} in {{Nonaqueous Solvents}}},
  author = {Marcus, Yizhak and Hefter, Glenn},
  year = 2004,
  month = jul,
  journal = {Chemical Reviews},
  volume = {104},
  number = {7},
  pages = {3405--3452},
  issn = {0009-2665, 1520-6890},
  doi = {10.1021/cr030047d},
  urldate = {2026-04-03},
  langid = {english}
}

@article{marcusThermodynamicsSolvationIons1991,
  title = {Thermodynamics of Solvation of Ions. {{Part}} 5.---{{Gibbs}} Free Energy of Hydration at 298.15 {{K}}},
  author = {Marcus, Yizhak},
  year = 1991,
  journal = {J. Chem. Soc., Faraday Trans.},
  volume = {87},
  number = {18},
  pages = {2995--2999},
  issn = {0956-5000, 1364-5455},
  doi = {10.1039/FT9918702995},
  urldate = {2026-10-18},
  langid = {english}
}

@article{marcusWaterStructureConcentrated2009,
  title = {On {{Water Structure}} in {{Concentrated Salt Solutions}}},
  author = {Marcus, Yizhak},
  year = 2009,
  month = may,
  journal = {Journal of Solution Chemistry},
  volume = {38},
  number = {5},
  pages = {513--516},
  issn = {0095-9782, 1572-8927},
  doi = {10.1007/s10953-009-9398-z},
  urldate = {2026-10-18},
  copyright = {http://www.springer.com/tdm},
  langid = {english}
}

@article{maribo-mogensenElectrolyteCPAEquation2015,
  title = {An Electrolyte {{CPA}} Equation of State for Mixed Solvent Electrolytes},
  author = {{Maribo-Mogensen}, Bj{\o}rn and Thomsen, Kaj and Kontogeorgis, Georgios M.},
  year = 2015,
  month = sep,
  journal = {AIChE Journal},
  volume = {61},
  number = {9},
  pages = {2933--2950},
  issn = {00011541},
  doi = {10.1002/aic.14829},
  urldate = {2017-03-12},
  langid = {english}
}

@article{maribo-mogensenModelingDielectricProperties2013a,
  title = {Modeling of {{Dielectric Properties}} of {{Aqueous Salt Solutions}} with an {{Equation}} of {{State}}},
  author = {{Maribo-Mogensen}, Bj{\o}rn and Kontogeorgis, Georgios M. and Thomsen, Kaj},
  year = 2013,
  month = sep,
  journal = {The Journal of Physical Chemistry B},
  volume = {117},
  number = {36},
  pages = {10523--10533},
  issn = {1520-6106, 1520-5207},
  doi = {10.1021/jp403375t},
  urldate = {2026-10-18},
  langid = {english}
}

@article{markgrafExtensionOpenCOSMORSFull2026,
  title = {Extension of {{openCOSMO-RS}} into a Full Open-Source Equation of State: {{Implementation}}, Parameterization, and Benchmarking},
  shorttitle = {Extension of {{openCOSMO-RS}} into a Full Open-Source Equation of State},
  author = {Markgraf, Jan and Lisboa Girardi, D. G. and Smirnova, Irina and M{\"u}ller, Simon},
  year = 2027,
  month = jan,
  journal = {Fluid Phase Equilibria},
  volume = {612},
  pages = {114820},
  issn = {03783812},
  doi = {10.1016/j.fluid.2026.114820},
  langid = {english}
}

@article{ahmedFastAccuratePrediction2026,
  title = {Fast and Accurate Prediction of Solid--Liquid Equilibria in Type {{V}} Eutectic Systems Using {{openCOSMO-RS}}},
  author = {Ahmed, Abdelkadr S. and M{\"u}ller, Simon and Fraterrigo Garofalo, Silvia and Tommasi, Tonia and Smirnova, Irina},
  year = 2026,
  month = may,
  journal = {Journal of Chemical \& Engineering Data},
  volume = {71},
  number = {6},
  pages = {2415--2424},
  issn = {0021-9568},
  doi = {10.1021/acs.jced.5c00865},
  langid = {english}
}

@article{grigorashComprehensiveApproachIncorporating2025,
  title = {A Comprehensive Approach to Incorporating Intermolecular Dispersion into the {{openCOSMO-RS}} Model. {{Part}} 1: {{Halocarbons}}},
  author = {Grigorash, Daria and M{\"u}ller, Simon and Paricaud, Patrice and Stenby, Erling H. and Smirnova, Irina and Yan, Wei},
  year = 2025,
  month = may,
  journal = {Chemical Engineering Science},
  volume = {309},
  pages = {121425},
  issn = {00092509},
  doi = {10.1016/j.ces.2025.121425},
  langid = {english}
}

@article{grigorashComprehensiveApproachIncorporating2026,
  title = {A Comprehensive Approach to Incorporating Intermolecular Dispersion into the {{openCOSMO-RS}} Model. {{Part}} 2: {{Atomic}} Polarizabilities},
  author = {Grigorash, Daria and M{\"u}ller, Simon and Heid, Esther and Neese, Frank and Liakos, Dimitrios G. and Riplinger, Christoph and {Garcia-Rat{\'e}s}, Miquel and Paricaud, Patrice and Stenby, Erling H. and Smirnova, Irina and Yan, Wei},
  year = 2026,
  month = jan,
  journal = {Chemical Engineering Science},
  volume = {319},
  pages = {122170},
  issn = {00092509},
  doi = {10.1016/j.ces.2025.122170},
  langid = {english}
}

@article{nevolianisSolvationFreeEnergies2025,
  title = {Solvation Free Energies of Anions: {{From}} Curated Reference Data to Predictive Models},
  author = {Nevolianis, Thomas and Zheng, Jonathan W. and M{\"u}ller, Simon and Baumann, Matthias and Tshepelevitsh, Sofja and Kaljurand, Ivari and Leito, Ivo and Smirnova, Irina and Green, William H. and Leonhard, Kai},
  year = 2025,
  month = aug,
  journal = {Journal of the American Chemical Society},
  volume = {147},
  number = {34},
  pages = {30626--30646},
  issn = {0002-7863},
  doi = {10.1021/jacs.5c02578},
  langid = {english}
}

@article{krogerPredictionSolvationFree2020,
  title = {Prediction of Solvation Free Energies of Ionic Solutes in Neutral Solvents},
  author = {Kr{\"o}ger, Leif C. and M{\"u}ller, Simon and Smirnova, Irina and Leonhard, Kai},
  year = 2020,
  month = may,
  journal = {The Journal of Physical Chemistry A},
  volume = {124},
  number = {20},
  pages = {4171--4181},
  issn = {1089-5639},
  doi = {10.1021/acs.jpca.0c01606},
  langid = {english}
}

@article{arradThermodynamicModelingUsing2024,
  title = {Thermodynamic Modeling Using Extended {{UNIQUAC}} and {{COSMO-RS-ES}} Models: {{Case}} Study of the Cesium Nitrate--Water System over a Large Range of Temperatures},
  author = {Arrad, Mouad and Thomsen, Kaj and M{\"u}ller, Simon and Smirnova, Irina},
  year = 2024,
  month = may,
  journal = {Fluid Phase Equilibria},
  volume = {580},
  pages = {114037},
  issn = {03783812},
  doi = {10.1016/j.fluid.2024.114037},
  langid = {english}
}

@dataset{mullerElectrolyteDatabase2026,
  title = {An Electrolyte Equation of State Based on {{openCOSMO-RS}}: {{From}} Aqueous Solutions to Non-Aqueous Systems},
  author = {M{\"u}ller, Simon},
  year = 2026,
  month = oct,
  publisher = {Zenodo},
  doi = {10.5281/zenodo.23668298},
  note = {Dataset}
}

@article{mullerHowCrackSMILES2025,
  title = {How to Crack a {{SMILES}}: Automatic Crosschecked Chemical Structure Resolution across Multiple Services Using {{MoleculeResolver}}},
  author = {M{\"u}ller, Simon},
  year = 2025,
  month = aug,
  journal = {Journal of Cheminformatics},
  volume = {17},
  number = {1},
  pages = {117},
  issn = {1758-2946},
  doi = {10.1186/s13321-025-01064-7},
  langid = {english}
}

@article{marquesImpactExtendedLongrange2023,
  title = {Impact of Extended Long-Range Electrostatics on the Correlation of Liquid-Liquid Equilibria in Aqueous Ionic Liquid Systems},
  author = {Marques, Hugo and Gonz{\'a}lez De Castilla, Andr{\'e}s and M{\"u}ller, Simon and Smirnova, Irina},
  year = 2023,
  month = jun,
  journal = {Fluid Phase Equilibria},
  volume = {569},
  pages = {113765},
  issn = {03783812},
  doi = {10.1016/j.fluid.2023.113765},
  urldate = {2026-03-05},
  langid = {english}
}

@article{mayJESSJointExpert2010,
  title = {{{JESS}}, a {{Joint Expert Speciation System}} -- {{IV}}: {{A}} Large Database of Aqueous Solution Physicochemical Properties with an Automatic Means of Achieving Thermodynamic Consistency},
  shorttitle = {{{JESS}}, a {{Joint Expert Speciation System}} -- {{IV}}},
  author = {May, Peter M. and Rowland, Darren and K{\"o}nigsberger, Erich and Hefter, Glenn},
  year = 2010,
  month = apr,
  journal = {Talanta},
  volume = {81},
  number = {1-2},
  pages = {142--148},
  issn = {00399140},
  doi = {10.1016/j.talanta.2009.11.049},
  urldate = {2026-10-19},
  copyright = {https://www.elsevier.com/tdm/userlicense/1.0/},
  langid = {english}
}

@article{milleroMolalVolumesElectrolytes1971a,
  title = {Molal Volumes of Electrolytes},
  author = {Millero, Frank J.},
  year = 1971,
  month = apr,
  journal = {Chemical Reviews},
  volume = {71},
  number = {2},
  pages = {147--176},
  issn = {0009-2665, 1520-6890},
  doi = {10.1021/cr60270a001},
  urldate = {2026-10-18},
  langid = {english}
}

@article{mohsRevisedLIQUACLIFAC2013,
  title = {A Revised {{LIQUAC}} and {{LIFAC}} Model ({{LIQUAC}}*/{{LIFAC}}*) for the Prediction of Properties of Electrolyte Containing Solutions},
  author = {Mohs, Andr{\'e} and Gmehling, J{\"u}rgen},
  year = 2013,
  month = jan,
  journal = {Fluid Phase Equilibria},
  volume = {337},
  pages = {311--322},
  issn = {03783812},
  doi = {10.1016/j.fluid.2012.09.023},
  urldate = {2026-07-26},
  copyright = {https://www.elsevier.com/tdm/userlicense/1.0/},
  langid = {english}
}

@phdthesis{muellerApplicationRefinementCOSMORSES2020,
  title = {{Application and refinement of COSMO-RS-ES for calculating phase equilibria of electrolyte systems at high concentrations in mixed and non-aqueous solvents}},
  author = {M{\"u}ller, Simon},
  year = 2020,
  month = aug,
  doi = {10.15480/882.2994},
  urldate = {2026-02-12},
  collaborator = {{TUHH Universit\"atsbibliothek} and Smirnova, Irina},
  copyright = {Creative Commons Attribution 4.0 International,},
  langid = {english},
  school = {Hamburg University of Technology},
  address = {Hamburg}
}

@article{mullerCalculationThermodynamicEquilibria2020,
  title = {Calculation of Thermodynamic Equilibria with the Predictive Electrolyte Model {{COSMO-RS-ES}}: {{Improvements}} for Low Permittivity Systems},
  shorttitle = {Calculation of Thermodynamic Equilibria with the Predictive Electrolyte Model {{COSMO-RS-ES}}},
  author = {M{\"u}ller, Simon and Gonz{\'a}lez De Castilla, Andr{\'e}s and Taeschler, Christoph and Klein, Andreas and Smirnova, Irina},
  year = 2020,
  month = feb,
  journal = {Fluid Phase Equilibria},
  volume = {506},
  pages = {112368},
  issn = {03783812},
  doi = {10.1016/j.fluid.2019.112368},
  urldate = {2026-03-05},
  langid = {english}
}

@article{mullerEvaluationRefinementNovel2019a,
  title = {Evaluation and Refinement of the Novel Predictive Electrolyte Model {{COSMO-RS-ES}} Based on Solid-Liquid Equilibria of Salts and {{Gibbs}} Free Energies of Transfer of Ions},
  author = {M{\"u}ller, Simon and Gonz{\'a}lez De Castilla, Andr{\'e}s and Taeschler, Christoph and Klein, Andreas and Smirnova, Irina},
  year = 2019,
  month = mar,
  journal = {Fluid Phase Equilibria},
  volume = {483},
  pages = {165--174},
  issn = {03783812},
  doi = {10.1016/j.fluid.2018.10.023},
  urldate = {2026-03-05},
  langid = {english}
}

@article{mullerPredictingSolvationFree2025,
  title = {Predicting Solvation Free Energies for Neutral Molecules in Any Solvent with {{openCOSMO-RS}}},
  author = {M{\"u}ller, Simon and Nevolianis, Thomas and {Garcia-Rat{\'e}s}, Miquel and Riplinger, Christoph and Leonhard, Kai and Smirnova, Irina},
  year = 2025,
  month = feb,
  journal = {Fluid Phase Equilibria},
  volume = {589},
  pages = {114250},
  issn = {03783812},
  doi = {10.1016/j.fluid.2024.114250},
  urldate = {2026-03-09},
  langid = {english}
}

@article{nguyenCorrelationsDensitiesAqueous2016a,
  title = {Correlations for {{Densities}} of {{Aqueous Electrolyte Solutions}}},
  author = {Nguyen, Nguyen H. and Hussain, Fazle and Chen, Chau-Chyun},
  year = 2016,
  month = feb,
  journal = {Journal of Chemical \& Engineering Data},
  volume = {61},
  number = {2},
  pages = {740--747},
  issn = {0021-9568, 1520-5134},
  doi = {10.1021/acs.jced.5b00491},
  urldate = {2026-08-13},
  langid = {english}
}

@article{novakExtensionESAFTVRMie2023,
  title = {Extension of the {{eSAFT-VR Mie}} Equation of State from Aqueous to Non-Aqueous Electrolyte Solutions},
  author = {Novak, Nefeli and Kontogeorgis, Georgios M. and Castier, Marcelo and Economou, Ioannis G.},
  year = 2023,
  month = feb,
  journal = {Fluid Phase Equilibria},
  volume = {565},
  pages = {113618},
  issn = {0378-3812},
  doi = {10.1016/j.fluid.2022.113618},
  urldate = {2026-08-13}
}

@article{novakMixedSolventElectrolyte2023,
  title = {Mixed {{Solvent Electrolyte Solutions}}: {{A Review}} and {{Calculations}} with the {{eSAFT-VR Mie Equation}} of {{State}}},
  shorttitle = {Mixed {{Solvent Electrolyte Solutions}}},
  author = {Novak, Nefeli and Kontogeorgis, Georgios M. and Castier, Marcelo and Economou, Ioannis G.},
  year = 2023,
  month = aug,
  journal = {Industrial \& Engineering Chemistry Research},
  volume = {62},
  number = {34},
  pages = {13646--13665},
  publisher = {American Chemical Society},
  issn = {0888-5885},
  doi = {10.1021/acs.iecr.3c00717},
  urldate = {2023-11-11}
}

@article{olsenComparisonsEquationState2023,
  title = {Comparisons of Equation of State Models for Electrolytes: {{e-CPA}} and {{e-PPC-SAFT}}},
  shorttitle = {Comparisons of Equation of State Models for Electrolytes},
  author = {Olsen, Martin Due and Kontogeorgis, Georgios M. and {de Hemptinne}, Jean-Charles and Liang, Xiaodong and {von Solms}, Nicolas},
  year = 2023,
  month = aug,
  journal = {Fluid Phase Equilibria},
  volume = {571},
  pages = {113804},
  issn = {0378-3812},
  doi = {10.1016/j.fluid.2023.113804},
  urldate = {2023-03-28},
  langid = {english}
}

@article{panayiotouStatisticalThermodynamicsRMer1982,
  title = {Statistical {{Thermodynamics}} of R-{{Mer Fluids}} and {{Their Mixtures}}},
  author = {Panayiotou, C and Vera, J H},
  year = 1982,
  month = sep,
  journal = {Polymer Journal},
  volume = {14},
  number = {9},
  pages = {681--694},
  issn = {0032-3896, 1349-0540},
  doi = {10.1295/polymj.14.681},
  urldate = {2026-07-26}
}

@article{paulingSIZESIONSSTRUCTURE1927a,
  title = {{{THE SIZES OF IONS AND THE STRUCTURE OF IONIC CRYSTALS}}},
  author = {Pauling, Linus},
  year = 1927,
  month = mar,
  journal = {Journal of the American Chemical Society},
  volume = {49},
  number = {3},
  pages = {765--790},
  issn = {0002-7863, 1520-5126},
  doi = {10.1021/ja01402a019},
  urldate = {2026-07-30},
  langid = {english}
}

@article{rashinReevaluationBornModel1985,
  title = {Reevaluation of the {{Born}} Model of Ion Hydration},
  author = {Rashin, Alexander A. and Honig, Barry},
  year = 1985,
  month = dec,
  journal = {The Journal of Physical Chemistry},
  volume = {89},
  number = {26},
  pages = {5588--5593},
  issn = {0022-3654, 1541-5740},
  doi = {10.1021/j100272a006},
  urldate = {2026-10-18},
  langid = {english}
}

@article{reynoldsStructureMakerstructureBreaker2024,
  title = {The Structure Maker-Structure Breaker Categories for Aqueous Electrolytes Signify Strength of {{Lewis}} Acidity/Basicity Relative to Water},
  author = {Reynolds, Jacob G.},
  year = 2024,
  month = feb,
  journal = {Journal of Molecular Liquids},
  volume = {395},
  pages = {123824},
  issn = {01677322},
  doi = {10.1016/j.molliq.2023.123824},
  urldate = {2026-10-25},
  langid = {english}
}

@article{ruebenPermittivityModelingElectrolyte2024,
  title = {Permittivity {{Modeling}} in {{Electrolyte PC-SAFT}}},
  author = {Rueben, Lisa and Rehner, Philipp and Gross, Joachim and Bardow, Andr{\'e}},
  year = 2024,
  month = sep,
  journal = {Journal of Chemical \& Engineering Data},
  volume = {69},
  number = {9},
  pages = {3044--3054},
  issn = {0021-9568, 1520-5134},
  doi = {10.1021/acs.jced.4c00347},
  urldate = {2026-03-13},
  copyright = {https://creativecommons.org/licenses/by/4.0/},
  langid = {english}
}

@article{ruebenPredictingRelativeStatic2024,
  title = {Predicting the {{Relative Static Permittivity}}: A {{Group Contribution Method Based}} on {{Perturbation Theory}}},
  shorttitle = {Predicting the {{Relative Static Permittivity}}},
  author = {Rueben, Lisa and Schilling, Johannes and Rehner, Philipp and M{\"u}ller, Simon and Esper, Timm and Bardow, Andr{\'e} and Gross, Joachim},
  year = 2024,
  month = feb,
  journal = {Journal of Chemical \& Engineering Data},
  volume = {69},
  number = {2},
  pages = {414--426},
  issn = {0021-9568, 1520-5134},
  doi = {10.1021/acs.jced.3c00323},
  urldate = {2026-03-13},
  copyright = {https://creativecommons.org/licenses/by/4.0/},
  langid = {english}
}

@article{sanchezElementaryMolecularTheory1976,
  title = {An Elementary Molecular Theory of Classical Fluids. {{Pure}} Fluids},
  author = {Sanchez, Isaac C. and Lacombe, Robert H.},
  year = 1976,
  month = oct,
  journal = {The Journal of Physical Chemistry},
  volume = {80},
  number = {21},
  pages = {2352--2362},
  issn = {0022-3654, 1541-5740},
  doi = {10.1021/j100562a008},
  urldate = {2026-07-26},
  langid = {english}
}

@article{sanderCalculationVapourliquidEquilibria1986,
  title = {Calculation of Vapour-Liquid Equilibria in Mixed Solvent/Salt Systems Using an Extended {{UNIQUAC}} Equation},
  author = {Sander, Bo and Fredenslund, Aage and Rasmussen, Peter},
  year = 1986,
  month = jan,
  journal = {Chemical Engineering Science},
  volume = {41},
  number = {5},
  pages = {1171--1183},
  issn = {0009-2509},
  doi = {10.1016/0009-2509(86)87090-7},
  urldate = {2026-07-26}
}

@article{schreckenbergModellingThermodynamicSolvation2014,
  title = {Modelling of the Thermodynamic and Solvation Properties of Electrolyte Solutions with the Statistical Associating Fluid Theory for Potentials of Variable Range},
  author = {Schreckenberg, Jens M.A. and Dufal, Simon and Haslam, Andrew J. and Adjiman, Claire S. and Jackson, George and Galindo, Amparo},
  year = 2014,
  month = sep,
  journal = {Molecular Physics},
  volume = {112},
  number = {17},
  pages = {2339--2364},
  issn = {0026-8976, 1362-3028},
  doi = {10.1080/00268976.2014.910316},
  urldate = {2015-04-19},
  langid = {english}
}

@article{selamThermodynamicModelStrong2018,
  title = {A Thermodynamic Model for Strong Aqueous Electrolytes Based on the {{eSAFT-VR Mie}} Equation of State},
  author = {Selam, Muaz A. and Economou, Ioannis G. and Castier, Marcelo},
  year = 2018,
  month = may,
  journal = {Fluid Phase Equilibria},
  volume = {464},
  pages = {47--63},
  issn = {0378-3812},
  doi = {10/gc4mz7},
  urldate = {2018-03-12}
}

@article{silvaHowAccountConcentration2023,
  title = {How to Account for the Concentration Dependency of Relative Permittivity in the {{Debye}}--{{H\"uckel}} and {{Born}} Equations},
  author = {Silva, Gabriel M. and Liang, Xiaodong and Kontogeorgis, Georgios M.},
  year = 2023,
  month = mar,
  journal = {Fluid Phase Equilibria},
  volume = {566},
  pages = {113671},
  issn = {03783812},
  doi = {10.1016/j.fluid.2022.113671},
  urldate = {2026-07-26},
  langid = {english}
}

@article{stokesIonicHydrationActivity1948,
  title = {Ionic {{Hydration}} and {{Activity}} in {{Electrolyte Solutions}}},
  author = {Stokes, R. H. and Robinson, R. A.},
  year = 1948,
  month = may,
  journal = {Journal of the American Chemical Society},
  volume = {70},
  number = {5},
  pages = {1870--1878},
  issn = {0002-7863, 1520-5126},
  doi = {10.1021/ja01185a065},
  urldate = {2026-10-18},
  langid = {english}
}

@article{thomsenModelingVaporLiquid1999,
  title = {Modeling of Vapor--Liquid--Solid Equilibrium in Gas--Aqueous Electrolyte Systems},
  author = {Thomsen, Kaj and Rasmussen, Peter},
  year = 1999,
  month = jun,
  journal = {Chemical Engineering Science},
  volume = {54},
  number = {12},
  pages = {1787--1802},
  issn = {0009-2509},
  doi = {10.1016/S0009-2509(99)00019-6},
  urldate = {2026-07-26}
}

@article{toureCOSMORSPDHSNewPredictive2014,
  title = {{{COSMO-RS-PDHS}}: {{A}} New Predictive Model for Aqueous Electrolytes Solutions},
  shorttitle = {{{COSMO-RS-PDHS}}},
  author = {Toure, Oumar and Audonnet, Fabrice and Lebert, Andr{\'e} and Dussap, Claude-Gilles},
  year = 2014,
  month = dec,
  journal = {Chemical Engineering Research and Design},
  volume = {92},
  number = {12},
  pages = {2873--2883},
  issn = {02638762},
  doi = {10.1016/j.cherd.2014.06.020},
  urldate = {2026-07-26},
  langid = {english}
}

@article{walkerImportanceRelativeStatic2022,
  title = {Importance of the {{Relative Static Permittivity}} in Electrolyte {{SAFT-VR Mie Equations}} of {{State}}},
  author = {Walker, Pierre J. and Liang, Xiaodong and Kontogeorgis, Georgios M.},
  year = 2022,
  month = jan,
  journal = {Fluid Phase Equilibria},
  volume = {551},
  pages = {113256},
  issn = {0378-3812},
  doi = {10.1016/j.fluid.2021.113256},
  urldate = {2026-08-13}
}

@article{wangComputationDielectricConstants2001,
  title = {Computation of Dielectric Constants of Solvent Mixtures and Electrolyte Solutions},
  author = {Wang, Peiming and Anderko, Andrzej},
  year = 2001,
  month = aug,
  journal = {Fluid Phase Equilibria},
  volume = {186},
  number = {1},
  pages = {103--122},
  issn = {0378-3812},
  doi = {10.1016/S0378-3812(01)00507-6},
  urldate = {2026-07-26}
}

@article{wangExtensionCOSMOSACSolvation2011,
  title = {Extension of {{COSMO-SAC Solvation Model}} for {{Electrolytes}}},
  author = {Wang, Shu and Song, Yuhua and Chen, Chau-Chyun},
  year = 2011,
  month = jan,
  journal = {Industrial \& Engineering Chemistry Research},
  volume = {50},
  number = {1},
  pages = {176--187},
  issn = {0888-5885, 1520-5045},
  doi = {10.1021/ie100689g},
  urldate = {2026-07-26},
  langid = {english}
}

@article{yangBenchmarkDatabaseMixedSolvent2022,
  title = {A {{Benchmark Database}} for {{Mixed-Solvent Electrolyte Solutions}}: {{Consistency Analysis Using E-NRTL}}},
  shorttitle = {A {{Benchmark Database}} for {{Mixed-Solvent Electrolyte Solutions}}},
  author = {Yang, Fufang and Ngo, Tri Dat and Kontogeorgis, Georgios M. and De Hemptinne, Jean-Charles},
  year = 2022,
  month = oct,
  journal = {Industrial \& Engineering Chemistry Research},
  volume = {61},
  number = {42},
  pages = {15576--15593},
  issn = {0888-5885, 1520-5045},
  doi = {10.1021/acs.iecr.2c00059},
  urldate = {2026-03-26},
  copyright = {https://doi.org/10.15223/policy-029},
  langid = {english}
}

@article{yangSystematicEvaluationParameterization2023,
  title = {Systematic Evaluation of Parameterization Approaches for the {{ePPC-SAFT}} Model for Aqueous Alkali Halide Solutions},
  author = {Yang, Fufang and Ngo, Tri Dat and Pinto, Juan Sebastian Roa and Kontogeorgis, Georgios M. and {de Hemptinne}, Jean-Charles},
  year = 2023,
  month = mar,
  journal = {Fluid Phase Equilibria},
  volume = {570},
  pages = {113778},
  issn = {0378-3812},
  doi = {10.1016/j.fluid.2023.113778},
  urldate = {2023-03-06},
  langid = {english}
}

@article{yangVaporLiquidEquilibriaWater1998,
  title = {{{Vapor}}-{{Liquid Equilibria}} of {{Water}} + {{Methanol}} in the {{Presence}} of {{Mixed Salts}}},
  author = {Yang, Sung-Oh and Lee, Chul Soo},
  year = 1998,
  month = jul,
  journal = {Journal of Chemical \& Engineering Data},
  volume = {43},
  number = {4},
  pages = {558--561},
  issn = {0021-9568, 1520-5134},
  doi = {10.1021/je970286w},
  urldate = {2026-10-25},
  langid = {english}
}

@article{yanPredictionVaporLiquid1999,
  title = {Prediction of Vapor--Liquid Equilibria in Mixed-Solvent Electrolyte Systems Using the Group Contribution Concept},
  author = {Yan, Weidong and Topphoff, Magnus and Rose, Christian and Gmehling, J{\"u}rgen},
  year = 1999,
  month = aug,
  journal = {Fluid Phase Equilibria},
  volume = {162},
  number = {1},
  pages = {97--113},
  issn = {0378-3812},
  doi = {10.1016/S0378-3812(99)00201-0},
  urldate = {2026-07-26}
}

@article{zavitsasPropertiesWaterSolutions2001,
  title = {Properties of {{Water Solutions}} of {{Electrolytes}} and {{Nonelectrolytes}}},
  author = {Zavitsas, Andreas A.},
  year = 2001,
  month = aug,
  journal = {The Journal of Physical Chemistry B},
  volume = {105},
  number = {32},
  pages = {7805--7817},
  issn = {1520-6106, 1520-5207},
  doi = {10.1021/jp011053l},
  urldate = {2026-10-18},
  langid = {english}
}

@article{zuendThermodynamicModelMixed2008,
  title = {A Thermodynamic Model of Mixed Organic-Inorganic Aerosols to Predict Activity Coefficients},
  author = {Zuend, A. and Marcolli, C. and Luo, B. P. and Peter, T.},
  year = 2008,
  month = aug,
  journal = {Atmospheric Chemistry and Physics},
  volume = {8},
  number = {16},
  pages = {4559--4593},
  issn = {1680-7324},
  doi = {10.5194/acp-8-4559-2008},
  urldate = {2026-07-26},
  copyright = {https://creativecommons.org/licenses/by/3.0/},
  langid = {english}
}

@article{wagnerIAPWSFormulation1995Thermodynamic2002,
  title = {The {{IAPWS}} Formulation 1995 for the Thermodynamic Properties of Ordinary Water Substance for General and Scientific Use},
  author = {Wagner, W. and Pru{\ss}, A.},
  year = 2002,
  journal = {Journal of Physical and Chemical Reference Data},
  volume = {31},
  number = {2},
  pages = {387--535},
  doi = {10.1063/1.1461829}
}

@article{barthelUntersuchungenDispersionKomplexen1970,
  title = {{{Untersuchungen}} zur {{Dispersion}} der komplexen {{Dielektrizit{\"a}tskonstante}} w{\"a}{\ss}riger und nichtw{\"a}{\ss}riger {{Elektrolytl{\"o}sungen}}},
  author = {Barthel, J. and Schmithals, F. and Behret, H.},
  year = {1970},
  month = sep,
  journal = {Zeitschrift f{\"u}r Physikalische Chemie},
  volume = {71},
  number = {1--3},
  pages = {115--131},
  doi = {10.1524/zpch.1970.71.1_3.115}
}

@article{barthelUntersuchungenDispersionKomplexen1977,
  title = {{{Untersuchungen}} zur {{Dispersion}} der komplexen {{Dielektrizit{\"a}tskonstante}} w{\"a}{\ss}riger und nichtw{\"a}{\ss}riger {{Elektrolytl{\"o}sungen}}},
  author = {Barthel, J. and Kr{\"u}ger, J. and Schollmeyer, E.},
  year = {1977},
  month = feb,
  journal = {Zeitschrift f{\"u}r Physikalische Chemie},
  volume = {104},
  number = {1--3},
  pages = {59--72},
  doi = {10.1524/zpch.1977.104.1-3.059}
}

@incollection{burnsSaltEffectVaporliquid1979,
  title = {Salt {{Effect}} in {{Vapor-Liquid Equilibrium}} at {{Fixed Liquid Composition}}},
  author = {Burns, John A. and Furter, William F.},
  booktitle = {Thermodynamic {{Behavior}} of {{Electrolytes}} in {{Mixed Solvents}}---{{II}}},
  editor = {Furter, William F.},
  year = {1979},
  series = {Advances in {{Chemistry}}},
  volume = {177},
  pages = {11--26},
  publisher = {American Chemical Society},
  address = {Washington, DC},
  doi = {10.1021/ba-1979-0177.ch002}
}

@article{sunMolecularThermodynamicsSalt1996,
  title = {Molecular Thermodynamics of Salt Effect in Vapor-Liquid Equilibrium: Calculation of Isobaric {{VLE}} Salt Effect Parameters for Ethanol-Water-1-1 Type Electrolytic Systems},
  author = {Sun, Renyi},
  year = {1996},
  month = aug,
  journal = {CIESC Journal (Huagong Xuebao)},
  volume = {47},
  number = {4},
  pages = {401--409},
  note = {In Chinese}
}

@article{segaKineticDielectricDecrement2015,
  title = {Kinetic Dielectric Decrement Revisited: Phenomenology of Finite Ion Concentrations},
  shorttitle = {Kinetic Dielectric Decrement Revisited},
  author = {Sega, Marcello and Kantorovich, Sofia and Arnold, Axel},
  year = 2015,
  journal = {Physical Chemistry Chemical Physics},
  volume = {17},
  number = {1},
  pages = {130--133},
  doi = {10.1039/C4CP04182H},
  langid = {english}
}

@article{klamt_cosmospace_2002,
	title = {{COSMOSPACE}: {Alternative} to conventional activity‐coefficient models},
	volume = {48},
	copyright = {http://onlinelibrary.wiley.com/termsAndConditions\#vor},
	issn = {0001-1541, 1547-5905},
	shorttitle = {{COSMOSPACE}},
	url = {https://aiche.onlinelibrary.wiley.com/doi/10.1002/aic.690481023},
	doi = {10.1002/aic.690481023},
	language = {en},
	number = {10},
	urldate = {2024-08-30},
	journal = {AIChE Journal},
	author = {Klamt, Andreas and Krooshof, Gerard J. P. and Taylor, Ross},
	month = oct,
	year = {2002},
	pages = {2332--2349},
}

@article{zini_improved_2021,
	title = {An improved dispersive contribution for the {COSMO}-{SAC}-{Phi} equation of state},
	volume = {534},
	issn = {03783812},
	url = {https://linkinghub.elsevier.com/retrieve/pii/S0378381221000042},
	doi = {10.1016/j.fluid.2021.112942},
	language = {en},
	urldate = {2024-07-08},
	journal = {Fluid Phase Equilibria},
	author = {Zini, Lucas P. and Staudt, Paula B. and Soares, Rafael De P.},
	month = apr,
	year = {2021},
	pages = {112942},
}

@article{klamt_cosmo-rs_2000,
	title = {{COSMO}-{RS}: a novel and efficient method for the a priori prediction of thermophysical data of liquids},
	volume = {172},
	copyright = {https://www.elsevier.com/tdm/userlicense/1.0/},
	issn = {03783812},
	shorttitle = {{COSMO}-{RS}},
	url = {https://linkinghub.elsevier.com/retrieve/pii/S0378381200003575},
	doi = {10.1016/S0378-3812(00)00357-5},
	language = {en},
	number = {1},
	urldate = {2024-07-08},
	journal = {Fluid Phase Equilibria},
	author = {Klamt, Andreas and Eckert, Frank},
	month = jul,
	year = {2000},
	pages = {43--72},
}

@article{nelder_simplex_1965,
  title   = {A Simplex Method for Function Minimization},
  volume  = {7, 4},
  doi     = {10.1093/comjnl/7.4.308},
  journal = {The Computer Journal},
  author  = {J. A. Nelder and R. Mead},
  year    = {1965},
  pages   = {308–313}
}
\end{document}